\documentclass[%
 reprint,
 superscriptaddress,
nofootinbib,
 amsmath,amssymb,
 aps,
floatfix,
]{revtex4-2}

\usepackage{amsmath, amssymb}
\usepackage{bbold}
\usepackage{bm}
\usepackage{booktabs}
\usepackage{braket}
\usepackage{dsfont}
\usepackage{enumerate}
\usepackage{float}
\usepackage{graphicx}
\usepackage{hyperref}
\usepackage{multirow}
\usepackage{slashed}
\usepackage{textcomp}
\usepackage{url}
\usepackage[figure]{hypcap}
\usepackage{physics}
\usepackage[space-before-unit, detect-weight, detect-display-math, detect-inline-weight=math, per-mode=fraction]{siunitx}
\usepackage{tikz}
\usepackage[T5,T1]{fontenc}

\graphicspath{ {./images/} }

\newcommand{\unaryminus}{\scalebox{0.5}[1.0]{\( - \)}}
\newcommand\brabarb{\scalebox{.3}{(}\raisebox{-2pt}[0pt][0pt]{$\unaryminus$}\scalebox{.3}{)}}
\usepackage{pifont}

\begin{document}



\title{All-flavor constraints on nonstandard neutrino interactions and\\
generalized matter potential with three years of IceCube DeepCore data}



\begin{abstract}
We report constraints on nonstandard neutrino interactions (NSI) from the observation of atmospheric neutrinos with IceCube, limiting all individual coupling strengths from a single dataset. Furthermore, IceCube is the first experiment to constrain flavor-violating and nonuniversal couplings simultaneously.
Hypothetical NSI are generically expected to arise due to the exchange of a new heavy mediator particle.
Neutrinos propagating in matter scatter off fermions in the forward direction with negligible momentum transfer. Hence the study of the matter effect on neutrinos propagating in the Earth is sensitive to NSI independently of the energy scale of new physics. We present constraints on NSI obtained with an all-flavor event sample of atmospheric neutrinos based on three years of IceCube DeepCore data. The analysis uses neutrinos arriving from all directions, with reconstructed energies between \SI{5.6}{\giga\electronvolt} and \SI{100}{\giga\electronvolt}. We report constraints on the individual NSI coupling strengths considered singly, allowing for complex phases in the case of flavor-violating couplings. This demonstrates that IceCube is sensitive to the full NSI flavor structure at a level competitive with limits from the global analysis of all other experiments.
In addition, we investigate a generalized matter potential, whose overall scale and flavor structure are also constrained. 
\end{abstract}


\affiliation{III. Physikalisches Institut, RWTH Aachen University, D-52056 Aachen, Germany}
\affiliation{Department of Physics, University of Adelaide, Adelaide, 5005, Australia}
\affiliation{Dept. of Physics and Astronomy, University of Alaska Anchorage, 3211 Providence Dr., Anchorage, AK 99508, USA}
\affiliation{Dept. of Physics, University of Texas at Arlington, 502 Yates St., Science Hall Rm 108, Box 19059, Arlington, TX 76019, USA}
\affiliation{CTSPS, Clark-Atlanta University, Atlanta, GA 30314, USA}
\affiliation{School of Physics and Center for Relativistic Astrophysics, Georgia Institute of Technology, Atlanta, GA 30332, USA}
\affiliation{Dept. of Physics, Southern University, Baton Rouge, LA 70813, USA}
\affiliation{Dept. of Physics, University of California, Berkeley, CA 94720, USA}
\affiliation{Lawrence Berkeley National Laboratory, Berkeley, CA 94720, USA}
\affiliation{Institut f{\"u}r Physik, Humboldt-Universit{\"a}t zu Berlin, D-12489 Berlin, Germany}
\affiliation{Fakult{\"a}t f{\"u}r Physik {\&} Astronomie, Ruhr-Universit{\"a}t Bochum, D-44780 Bochum, Germany}
\affiliation{Universit{\'e} Libre de Bruxelles, Science Faculty CP230, B-1050 Brussels, Belgium}
\affiliation{Vrije Universiteit Brussel (VUB), Dienst ELEM, B-1050 Brussels, Belgium}
\affiliation{Department of Physics and Laboratory for Particle Physics and Cosmology, Harvard University, Cambridge, MA 02138, USA}
\affiliation{Dept. of Physics, Massachusetts Institute of Technology, Cambridge, MA 02139, USA}
\affiliation{Dept. of Physics and Institute for Global Prominent Research, Chiba University, Chiba 263-8522, Japan}
\affiliation{Department of Physics, Loyola University Chicago, Chicago, IL 60660, USA}
\affiliation{Dept. of Physics and Astronomy, University of Canterbury, Private Bag 4800, Christchurch, New Zealand}
\affiliation{Dept. of Physics, University of Maryland, College Park, MD 20742, USA}
\affiliation{Dept. of Astronomy, Ohio State University, Columbus, OH 43210, USA}
\affiliation{Dept. of Physics and Center for Cosmology and Astro-Particle Physics, Ohio State University, Columbus, OH 43210, USA}
\affiliation{Niels Bohr Institute, University of Copenhagen, DK-2100 Copenhagen, Denmark}
\affiliation{Dept. of Physics, TU Dortmund University, D-44221 Dortmund, Germany}
\affiliation{Dept. of Physics and Astronomy, Michigan State University, East Lansing, MI 48824, USA}
\affiliation{Dept. of Physics, University of Alberta, Edmonton, Alberta, Canada T6G 2E1}
\affiliation{Erlangen Centre for Astroparticle Physics, Friedrich-Alexander-Universit{\"a}t Erlangen-N{\"u}rnberg, D-91058 Erlangen, Germany}
\affiliation{Physik-department, Technische Universit{\"a}t M{\"u}nchen, D-85748 Garching, Germany}
\affiliation{D{\'e}partement de physique nucl{\'e}aire et corpusculaire, Universit{\'e} de Gen{\`e}ve, CH-1211 Gen{\`e}ve, Switzerland}
\affiliation{Dept. of Physics and Astronomy, University of Gent, B-9000 Gent, Belgium}
\affiliation{Dept. of Physics and Astronomy, University of California, Irvine, CA 92697, USA}
\affiliation{Karlsruhe Institute of Technology, Institute for Astroparticle Physics, D-76021 Karlsruhe, Germany }
\affiliation{Karlsruhe Institute of Technology, Institute of Experimental Particle Physics, D-76021 Karlsruhe, Germany }
\affiliation{Dept. of Physics and Astronomy, University of Kansas, Lawrence, KS 66045, USA}
\affiliation{SNOLAB, 1039 Regional Road 24, Creighton Mine 9, Lively, ON, Canada P3Y 1N2}
\affiliation{Department of Physics and Astronomy, UCLA, Los Angeles, CA 90095, USA}
\affiliation{Department of Physics, Mercer University, Macon, GA 31207-0001, USA}
\affiliation{Dept. of Astronomy, University of Wisconsin{\textendash}Madison, Madison, WI 53706, USA}
\affiliation{Dept. of Physics and Wisconsin IceCube Particle Astrophysics Center, University of Wisconsin{\textendash}Madison, Madison, WI 53706, USA}
\affiliation{Institute of Physics, University of Mainz, Staudinger Weg 7, D-55099 Mainz, Germany}
\affiliation{Department of Physics, Marquette University, Milwaukee, WI, 53201, USA}
\affiliation{Institut f{\"u}r Kernphysik, Westf{\"a}lische Wilhelms-Universit{\"a}t M{\"u}nster, D-48149 M{\"u}nster, Germany}
\affiliation{Bartol Research Institute and Dept. of Physics and Astronomy, University of Delaware, Newark, DE 19716, USA}
\affiliation{Dept. of Physics, Yale University, New Haven, CT 06520, USA}
\affiliation{Dept. of Physics, University of Oxford, Parks Road, Oxford OX1 3PU, UK}
\affiliation{Dept. of Physics, Drexel University, 3141 Chestnut Street, Philadelphia, PA 19104, USA}
\affiliation{Physics Department, South Dakota School of Mines and Technology, Rapid City, SD 57701, USA}
\affiliation{Dept. of Physics, University of Wisconsin, River Falls, WI 54022, USA}
\affiliation{Dept. of Physics and Astronomy, University of Rochester, Rochester, NY 14627, USA}
\affiliation{Department of Physics and Astronomy, University of Utah, Salt Lake City, UT 84112, USA}
\affiliation{Oskar Klein Centre and Dept. of Physics, Stockholm University, SE-10691 Stockholm, Sweden}
\affiliation{Dept. of Physics and Astronomy, Stony Brook University, Stony Brook, NY 11794-3800, USA}
\affiliation{Dept. of Physics, Sungkyunkwan University, Suwon 16419, Korea}
\affiliation{Institute of Basic Science, Sungkyunkwan University, Suwon 16419, Korea}
\affiliation{Dept. of Physics and Astronomy, University of Alabama, Tuscaloosa, AL 35487, USA}
\affiliation{Dept. of Astronomy and Astrophysics, Pennsylvania State University, University Park, PA 16802, USA}
\affiliation{Dept. of Physics, Pennsylvania State University, University Park, PA 16802, USA}
\affiliation{Dept. of Physics and Astronomy, Uppsala University, Box 516, S-75120 Uppsala, Sweden}
\affiliation{Dept. of Physics, University of Wuppertal, D-42119 Wuppertal, Germany}
\affiliation{DESY, D-15738 Zeuthen, Germany}

\author{R. Abbasi}
\affiliation{Department of Physics, Loyola University Chicago, Chicago, IL 60660, USA}
\author{M. Ackermann}
\affiliation{DESY, D-15738 Zeuthen, Germany}
\author{J. Adams}
\affiliation{Dept. of Physics and Astronomy, University of Canterbury, Private Bag 4800, Christchurch, New Zealand}
\author{J. A. Aguilar}
\affiliation{Universit{\'e} Libre de Bruxelles, Science Faculty CP230, B-1050 Brussels, Belgium}
\author{M. Ahlers}
\affiliation{Niels Bohr Institute, University of Copenhagen, DK-2100 Copenhagen, Denmark}
\author{M. Ahrens}
\affiliation{Oskar Klein Centre and Dept. of Physics, Stockholm University, SE-10691 Stockholm, Sweden}
\author{C. Alispach}
\affiliation{D{\'e}partement de physique nucl{\'e}aire et corpusculaire, Universit{\'e} de Gen{\`e}ve, CH-1211 Gen{\`e}ve, Switzerland}
\author{A. A. Alves Jr.}
\affiliation{Karlsruhe Institute of Technology, Institute for Astroparticle Physics, D-76021 Karlsruhe, Germany }
\author{N. M. Amin}
\affiliation{Bartol Research Institute and Dept. of Physics and Astronomy, University of Delaware, Newark, DE 19716, USA}
\author{R. An}
\affiliation{Department of Physics and Laboratory for Particle Physics and Cosmology, Harvard University, Cambridge, MA 02138, USA}
% ...
\date{\today}

\collaboration{IceCube Collaboration}\thanks{analysis@icecube.wisc.ed}
\noaffiliation

\maketitle

\tableofcontents


\section{\label{sec:intro}Introduction}
The collective evidence for flavor transitions of neutrinos propagating in vacuum and various types of matter conclusively demonstrates that at least two of the three known active neutrinos have mass~\cite{deSalas:2017kay,Capozzi:2018ubv,Nufit4.0,nufit4.1url}. Since these transitions typically exhibit an oscillatory pattern in the ratio of neutrino propagation distance to energy, they are also referred to as ``neutrino oscillations''~\cite{Pontecorvo:1957cp,Pontecorvo:1957qd}.
Since the neutrino mass scale is many orders of magnitude smaller than that of charged fermions, qualitatively different mechanisms for generating neutrino masses than in the Standard Model (SM) have been proposed~\cite{Gouvea:2016shl}.

Neutrino mass can be parametrized by viewing the SM as an effective theory~\cite{Weinberg:1967tq}, resulting from new physics beyond the SM (BSM) at some characteristic energy scale $\Lambda$ (typically above the energy scale of electroweak symmetry breaking; see reviews in Refs.~\cite{Langacker:1980js,Costa:1985vk}). The SM Lagrangian is then extended by nonrenormalizable operators of increasing energy dimension~\cite{Weinberg:1979sa}, with BSM neutrino interactions expected to arise at dimension six~\cite{PhysRevD.79.013007,Bischer:2019ttk} in a variety of models~\cite{Barbier:2004ez,Forero:2011pc,Boucenna:2014zba,Heeck:2018nzc,Babu:2019mfe}. Nonstandard neutrino interactions (NSI) are commonly understood to constitute the subset of neutral current (NC) and charged current (CC) operators that involve left-chiral neutrinos and both left- and right-chiral charged fermions (for a recent review, see~\cite{Farzan:2017xzy}), with the following effective Lagrangians~\cite{Wolfenstein:1978ue,PhysRevD.44.R935,Guzzo:1991hi,PhysRevD.44.1629,Grossman:1995wx}:

\begin{align}
	\mathcal{L}^\mathrm{NC}_\mathrm{NSI} &= -2\sqrt{2}G_F \epsilon^{f C}_{\alpha\beta} \left(\overline{\nu_\alpha}\gamma_\mu P_L\nu_\beta\right)\left(\overline{f}\gamma^\mu P_C f\right)\text{ ,}\label{eq:NC dim-6 NSI}\\
    \mathcal{L}^\mathrm{CC}_\mathrm{NSI} &= -2\sqrt{2}G_F \epsilon^{ff^\prime C}_{\alpha\beta} \left(\overline{\ell_\alpha}\gamma_\mu P_L\nu_\beta\right)\left(\overline{f^\prime}\gamma^\mu P_C f\right)\text{ .}\label{eq:CC dim-6 NSI}
\end{align}
Here, sums over the leptonic flavor indices $\alpha,\beta=e,\mu,\tau$ as well as the charged fermion type $f \neq f^\prime$ and chirality $C=L,R$ are implied, with $P$ denoting the chiral projector. In ordinary matter, $f,f^\prime=e,u,d$ indicate charged leptons and quarks of the first generation. The coefficients $\epsilon^{fC}_{\alpha\beta}$ and $\epsilon^{ff^\prime C}_{\alpha\beta}$ are the effective NC and CC NSI coupling strengths, respectively, normalized to Fermi's coupling constant $G_F$~\cite{Wilson:1968pwx}. Hence, the SM is recovered in the limit $\epsilon_{\alpha\beta} \to 0$.  The only NSI coupling strengths relevant to neutrinos propagating in matter with negligible incoherent interactions are given by
\begin{equation}
    \epsilon^{f}_{\alpha\beta} \equiv \epsilon^{fL}_{\alpha\beta} + \epsilon^{fR}_{\alpha\beta}\text{ .}\label{eq:NC NSI Lagrangian-level couplings}
\end{equation}
NSI couplings with $\alpha \neq \beta$ represent new sources of flavor violation (FV), whereas those with $\alpha = \beta$ accommodate new flavor-diagonal interactions, which could give rise to flavor nonuniversality (NU).

What sets neutrino oscillation experiments apart from other experiments is their unique capability to probe BSM scenarios responsible for NSI independently of the new physics energy scale $\Lambda$~\cite{Farzan:2017xzy}. Detailed global analyses of available neutrino oscillation data (e.g.~\cite{PhysRevD.95.072006,PhysRevLett.110.251801,PhysRevD.91.072004,PhysRevC.88.025501}) allowing for NSI have so far shown no statistically significant evidence for BSM interactions and have thus been used to place limits on NSI in a model-independent manner~\cite{Esteban:2018ppq}. Coupling strengths up to $\epsilon \sim \order{0.1}$ at a \SI{90}{\percent} confidence level continue to be allowed.

In this paper, we present a new search for NC NSI using atmospheric neutrinos\footnote{We denote both neutrinos and antineutrinos as ``neutrinos,'' unless a distinction is necessary.} interacting in the IceCube DeepCore detector.
Matter effects are expected for these neutrino trajectories, since their oscillation baselines range up to the order of the diameter of the Earth.
Compared to our previous study~\cite{Aartsen:2017xtt}, the present analysis is based on an extended event selection that includes neutrinos of all flavors, with reconstructed energies reaching up to \SI{100}{\giga\electronvolt}.
To obtain a high purity sample, the overwhelming background of atmospheric muons is reduced by approximately eight orders of magnitude through a series of containment and quality selection criteria~\cite{Aartsen:2019tjl}. 
Furthermore, whereas our earlier analysis constrained real-valued flavor-violating NSI in the $\mu$-$\tau$ sector via the disappearance of atmospheric muon neutrinos, we now constrain multiple, potentially complex-valued, NSI couplings, each through its simultaneous effects in all oscillation channels. In addition, we test a more general NSI flavor structure within a $CP$-conserving framework, proceeding in close analogy to the analysis of atmospheric and long-baseline accelerator neutrino experiments in recent global NSI fits~\cite{GonzalezGarcia:2011my,Gonzalez-Garcia:2013usa,Esteban:2018ppq}.
In a rigorous statistical approach, NSI hypotheses are tested by comparing Monte Carlo (MC) expectation to observation.

\section{\label{sec:flavour transitions}Neutrino flavor transitions in Earth matter with nonstandard interactions}
\subsection{\label{subsec:evolution}Evolution equation}

While the flavor evolution of ultrarelativistic neutrinos in vacuum depends solely on neutrino energy, mass-squared differences and the leptonic mixing matrix~\cite{Pontecorvo:1957cp,Pontecorvo:1957qd,Maki:1962mu,Tanabashi:2018oca}, evolution in matter introduces a potential position dependence~\cite{Cardall:1999bz,Akhmedov:2012mk}.
The energy-independent interaction Hamiltonian\footnote{While for neutrinos, the overall evolution is governed by the sum of the vacuum and matter Hamiltonian $H_{\nu}(x) = \left[H_\mathrm{vac} + H_\mathrm{mat}(x)\right]$, the evolution of antineutrinos follows $H_{\bar{\nu}}(x) = \left[H_\mathrm{vac} - H_\mathrm{mat}(x)\right]^*$~\cite{GonzalezGarcia:2011my}.} is identified with the matrix of effective neutrino potentials for coherent forward scattering in the flavor basis. For SM interactions in an unpolarized medium, the interaction Hamiltonian is
\begin{equation}
    H_\mathrm{mat}(x) = V(x) = V_\mathrm{CC}(x)\begin{pmatrix}
                 1 & 0 & 0\\ 
				 0 & 0 & 0\\
				 0 & 0 & 0\end{pmatrix}\text{ ,}
    \label{eq:Hmat}
\end{equation}
with the standard matter potential $V_\mathrm{CC}(x) = \sqrt{2}G_FN_e(x)$~\cite{Opher:1974drq,PhysRevD.27.1228}, where $N_e(x)$ is the local electron number density.\footnote{The corresponding Hamiltonian can be cast into the necessary NC form by means of a Fierz transformation~\cite{Fierz}.} This potential is responsible for the matter effects on neutrino propagation in the Sun and the Earth: the Mikheyev-Smirnov-Wolfenstein (MSW) effect and resonance~\cite{Wolfenstein:1978ue,Mikheyev1986} and, in the case of the Earth, parametric enhancement~\cite{Akhmedov:1998ui,Akhmedov:1988kd,Petcov:1998su}. See, e.g.,~\cite{Blennow:2013rca} for a review.

NSI contributions lead to a straightforward generalization of Eq.~(\ref{eq:Hmat}). Accounting for the hermiticity of the Hamiltonian, $H_\mathrm{mat}$ can be well approximated by
\begin{align}
    H_\mathrm{mat}(x) = V_\mathrm{CC}(x)\begin{pmatrix}
				1 + \epsilon^\oplus_{ee} - \epsilon^\oplus_{\mu\mu} & \epsilon^\oplus_{e\mu} & \epsilon^\oplus_{e\tau}\\ 
				 \epsilon^{\oplus*}_{e\mu} & 0 & \epsilon^\oplus_{\mu\tau}\\
				 \epsilon^{\oplus*}_{e\tau} & \epsilon^{\oplus*}_{\mu\tau} & \epsilon^\oplus_{\tau\tau} - \epsilon^\oplus_{\mu\mu}\end{pmatrix}\text{ ,}
	\label{eq:generalised Hmat final}
\end{align}
where a term proportional to $\epsilon^\oplus_{\mu\mu}\cdot\mathbb{1}$ was subtracted to reduce the dimensionality without observable consequences. This interaction Hamiltonian makes use of the constant effective NSI couplings, 
\begin{equation}
	\epsilon^\oplus_{\alpha\beta} \approx \epsilon^{e}_{\alpha\beta} + \epsilon^{p}_{\alpha\beta} + Y^\oplus_n \epsilon^{n}_{\alpha\beta}\text{ ,}
	\label{eq:constant Earth effective couplings protons neutrons}
\end{equation}
to electrons, protons and neutrons ($e$, $p$, $n$) in Earth matter, the latter including the nearly constant relative neutron-to-electron number density of the Earth, $Y^\oplus_n \equiv \expval{N_n(x)/N_e(x)} \approx 1.051$~\cite{Esteban:2018ppq}. The Hamiltonian is described by eight real NSI parameters (five amplitudes and three phases):
\begin{align}
    \epsilon^{\oplus}_{\alpha\alpha} - \epsilon^{\oplus}_{\mu\mu} &= \Re{ \epsilon^{\oplus}_{\alpha\alpha} - \epsilon^{\oplus}_{\mu\mu}}&\quad(\alpha\in{e,\tau})\text{ ,}\label{eq:real NU couplings}\\
    \big|\epsilon^\oplus_{\alpha\beta}\big|e^{i\delta_{\alpha\beta}} &= \epsilon^\oplus_{\alpha\beta}&\quad(\alpha\neq \beta)\text{ .}\label{eq:complex FC couplings}
\end{align}

This will in the following be referred to as ``standard parametrization''.

\subsection{\label{subsec:GMP}Generalized matter potential}
In addition to the above description of NSI, our analysis is carried out in the alternative parametrization~\cite{GonzalezGarcia:2011my,Fornengo:2001pm}
\begin{equation}
	H_\mathrm{mat}(x) = Q_\mathrm{rel}U_\mathrm{mat}D_\mathrm{mat}(x)U^\dagger_\mathrm{mat}Q^\dagger_\mathrm{rel}\text{ ,}
	\label{eq:vacuum Hamiltonian like parameterisation}
\end{equation}
with the three matrices on the right side defined as~\cite{Esteban:2018ppq}
\begin{align}
    D_\mathrm{mat}(x) &= V_\mathrm{CC}(x) \mathrm{diag}(\epsilon_\oplus, \epsilon^\prime_\oplus, 0)\label{eq:matter potential eigenvalues}\text{ ,}\\
    U_\mathrm{mat} &= R_{12}(\varphi_{12})R_{13}(\varphi_{13})\tilde{R}_{23}(\varphi_{23},\delta_\mathrm{NS})\label{eq:matter potential rotation}\text{ ,}\\
    Q_\mathrm{rel} &= \mathrm{diag}\left(e^{i\alpha_1}, e^{i\alpha_2}, e^{-i(\alpha_1 + \alpha_2)}\right)\label{eq:relative matter-vacuum rephasing}\text{ .}
\end{align}
Here $R_{12}(\varphi_{12})$ and $R_{13}(\varphi_{13})$ correspond to real rotations through the angles $\varphi_{12}$ and $\varphi_{13}$ in the $1$-$2$ and $1$-$3$ planes, respectively, whereas $\tilde{R}_{23}(\varphi_{23},\delta_\mathrm{NS})$ denotes a complex rotation through the angle $\varphi_{23}$ and the phase $\delta_\mathrm{NS}$. 

Since IceCube DeepCore is sensitive mainly to muon neutrino disappearance and existing data from atmospheric neutrino experiments has little sensitivity to $CP$-violating effects~\cite{Esteban:2018ppq}, the dimensionality of this parametrization can be reduced while approximately retaining model independence. We set $\epsilon^\prime_\oplus=0$~\cite{Friedland:2004ah,Friedland:2005vy}, set the phases $\alpha_{1,2} = 0$, and disregard $\varphi_{23}$ and $\delta_\mathrm{NS}$ as unphysical~\cite{GonzalezGarcia:2011my,Gonzalez-Garcia:2013usa} (see Appendix~\ref{app:detailed nsi params} for a complete justification). As a result, $H_\mathrm{mat}$ is real-valued and has three free parameters:%
\begin{equation}\label{eq:gmp_pars}
  \epsilon_{\bigoplus}\text{, }\varphi_{12}\text{, }\varphi_{13}\text{ .}
\end{equation}%
Any given point in the three-dimensional parameter space of this ``generalized matter potential'' (GMP) uniquely corresponds to a point in the standard parametrization described in Sec.~\ref{subsec:evolution}.
When the vacuum Hamiltonian $H_\mathrm{vac}$ is included in the $CP$-conserving framework by setting $\delta_\mathrm{CP}=0$, we can retain the usual minimal parameter ranges for the standard Pontecorvo-Maki-Nakagawa-Sakata (PMNS)~\cite{Pontecorvo:1957cp,Maki:1962mu} mixing parameters and neutrino mass-squared differences~\cite{deGouvea:2008nm} by choosing the ranges of the matter potential rotation angles as $-\pi/2 \leq \varphi_{ij} \leq \pi/2$.
Due to the generalized mass ordering degeneracy explained in detail in Appendix~\ref{app:detailed nsi params}, when $H_\mathrm{mat}$ is only described by $(\epsilon_\oplus, \varphi_{12}, \varphi_{13})$, it is sufficient to restrict $\Delta m^2_{31(32)} > 0$ and test both signs of $\epsilon_\oplus$. The two choices $(\epsilon_\oplus = \pm 1, \varphi_{12} = 0, \varphi_{13} = 0)$ correspond to neutrino propagation with SM interactions given the normal neutrino mass ordering (``$+$'') and the inverted neutrino mass ordering (``$-$''), respectively.

\subsection{\label{subsec:NSI in oscillation probs}NSI effects on the oscillation probability}
In our calculation we assume an atmospheric neutrino production height of $h=\SI{20}{\kilo\metre}$ above the surface~\cite{Honda:2015fha}. 
The zenith angle $\vartheta$ then geometrically fixes the oscillation baseline~\cite{Giunti:2007ry} ranging from ``upgoing,'' Earth-crossing ($\cos\vartheta = -1$, $d \approx \SI{1.3e4}{\kilo\metre}$) trajectories to ``downgoing'' ($\cos\vartheta = 1$, $d \approx \SI{20}{\kilo\metre}$) trajectories.

We approximate the Earth's matter density profile using twelve concentric uniform-density layers adopted from the preliminary reference Earth model (PREM)~\cite{prem}, with matter densities between about \SI[per-mode=symbol]{3}{\gram\per\centi\metre\cubed} and \SI[per-mode=symbol]{13}{\gram\per\centi\metre\cubed}.
We take the relative electron-to-nucleon number density $Y^c_e = 0.466$ for the Earth's inner and outer core; for the mantle we choose $Y^m_e = 0.496$. The (nominal) values for the PMNS mixing parameters and the neutrino mass-squared differences are taken from a global fit to neutrino oscillation data~\cite{nufit3,nufit3.2url}, except for $\delta_{CP} = 0$, to which the analysis is insensitive. 

In Fig.~\ref{fig:SM and NSI selected osc. probs 1d eps_oplus eps_tautau  eps_mutau} the oscillation probability $P_{\mu\tau}$ is shown for three different NSI parameters as a function of the neutrino energy $\SI{2}{\giga\electronvolt} \leq E_\nu \leq \SI{1000}{\giga\electronvolt}$ for an inclined trajectory that only crosses the Earth's mantle. The chosen zenith angle of $\cos(\vartheta) = -0.75$ corresponds to a baseline $L \approx \SI{9.6e3}{\kilo\metre}$. 
We show the corresponding standard interactions (SI) oscillation probability as a reference in each figure. Approximations employed in the discussions below are just for illustrative purposes. All oscillation probabilities underlying the analysis in this paper are obtained by solving the full three-neutrino evolution equation~\cite{PISA}.

\begin{figure}[h!t]
\centering
\includegraphics[width=\linewidth]{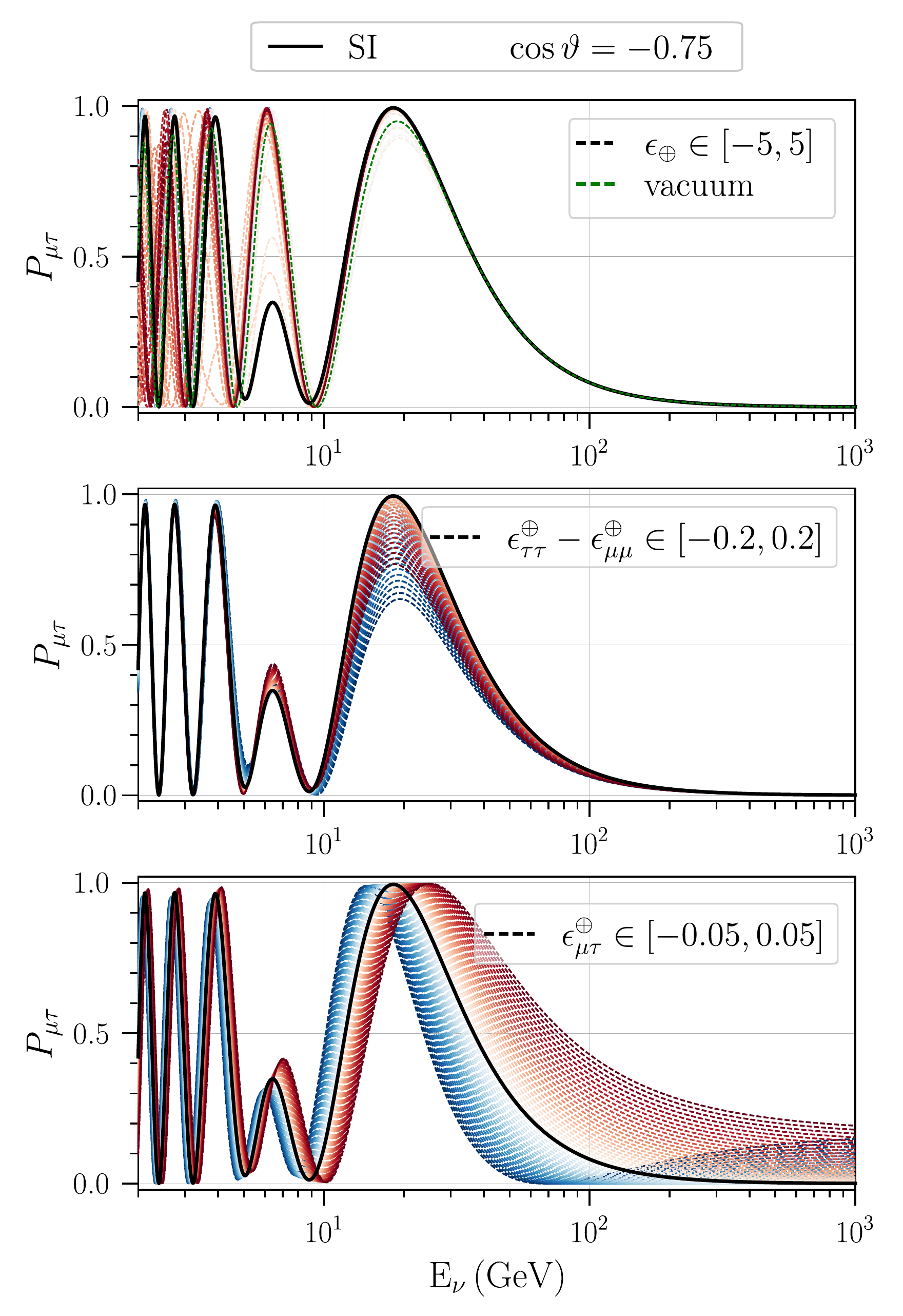}
\caption{\label{fig:SM and NSI selected osc. probs 1d eps_oplus eps_tautau eps_mutau}$P_{\mu\tau}$ oscillation probability of atmospheric neutrinos crossing the Earth at a zenith angle of $\cos(\vartheta)=-0.75$ vs. the neutrino energy $E_\nu$. Shown are different realizations of three NSI parameters, each varied separately. These exhibit the most prominent features and discrepancies from the SM interactions (SI) case, taking into account the different importance of individual channels.
In each panel, the SI case is represented by the black line. Blue dashed lines show the probabilities obtained for negative parameter values, while the red dashed lines are for positive values. Darker colors represent larger absolute values of the respective NSI parameter.
Top panel: The effective matter potential $\epsilon_\oplus$ of the GMP parametrization (cf. Sec.~\ref{subsec:GMP}) with $-5 \leq \epsilon_\oplus \leq 5$, restricting the matter potential to the $ee$ matrix element, yielding $\epsilon_\oplus = 1 + \epsilon^\oplus_{ee} - \epsilon^\oplus_{\mu\mu}$. Apart from the SI case ($\epsilon_\oplus=1$), the no interactions case (vacuum, $\epsilon_\oplus = 0$) is highlighted as a dashed green line. The blue lines denoting negative values are mostly covered by the dark red lines.
Center panel: The NSI nonuniversality strength $\epsilon^\oplus_{\tau\tau} - \epsilon^\oplus_{\mu\mu}$, with $-0.20 \leq \epsilon^\oplus_{\tau\tau} - \epsilon^\oplus_{\mu\mu} \leq 0.20$. 
Bottom panel: The NSI coupling strength $\epsilon^\oplus_{\mu\tau}$, with $-0.05 \leq \epsilon^\oplus_{\mu\tau} \leq 0.05$ and $\delta_{\mu\tau} = \SI{0}{\degree}$.
 }
\end{figure}

The top panel of Fig.~\ref{fig:SM and NSI selected osc. probs 1d eps_oplus eps_tautau eps_mutau} shows the oscillation probability $P_{\mu\tau}$ that results from varying the GMP parameter $\epsilon_\oplus$ while restricting the matter potential to the $ee$ matrix element, i.e., $\varphi_{ij}=0$. All of the cases correspond to a rescaling of the SM matter potential by the factor $V_\mathrm{CC}(x) \to V^\prime(x) = \epsilon_\oplus V_\mathrm{CC}(x) = \left(1 + \epsilon^\oplus_{ee} - \epsilon^\oplus_{\mu\mu}\right)V_\mathrm{CC}(x)$. 

The center and bottom panels of Fig.~\ref{fig:SM and NSI selected osc. probs 1d eps_oplus eps_tautau eps_mutau} show $P_{\mu\tau}$ when the only source of NSI is $\epsilon^\oplus_{\tau\tau} - \epsilon^\oplus_{\mu\mu}$ and $\epsilon^\oplus_{\mu\tau}$, respectively, in the latter assuming a real-valued NSI coupling for simplicity.
For the more general case of complex-valued $\epsilon^\oplus_{\mu\tau}$, cf. Eq.~(\ref{eq:complex FC couplings}), the value of the complex phase $\delta_{\mu\tau}$ affects the impact of the magnitude $\abs{\epsilon^\oplus_{\mu\tau}}$ on the oscillation probabilities in the $\mu$-$\tau$ sector. For example, their leading-order perturbative expansions~\cite{PhysRevD.77.013007} reveal that a purely imaginary coupling $\epsilon^\oplus_{\mu\tau} = i\abs{\epsilon^\oplus_{\mu\tau}}$ (corresponding to $\delta_{\mu\tau} = \SI{90}{\degree}, \SI{270}{\degree}$) is expected to result in less sensitivity at the probability level.

All oscillation probabilities $P_{\alpha\beta}$ that are not given in Fig.~\ref{fig:SM and NSI selected osc. probs 1d eps_oplus eps_tautau eps_mutau} show similar or subdominant effects and can be found in Appendix~\ref{app:phenomenology nsi at prob lvl} alongside a more detailed phenomenological discussion.

\section{\label{sec:DeepCore event selection}Event Selection with IceCube DeepCore}

The in-ice array of the IceCube Neutrino Observatory, located at the Geographic South Pole, consists of 5160 individual photosensors. These form a cubic-kilometer-volume detector for Cherenkov emission of charged particles propagating through the Antarctic ice shield~\cite{Aartsen:2016nxy}, allowing for the detection of neutrino interactions. The subarray at the center of IceCube, DeepCore, has an effective mass of approximately $\SI{10}{Mton}$ and is instrumented approximately five times more densely with respect to the standard IceCube array in order to lower IceCube's neutrino detection energy threshold to a few \si{\giga\electronvolt}~\cite{Collaboration:2011ym}. In event topologies, we differentiate between ``tracklike'' extended light depositions along muon trajectories caused by $\nu_\mu$ CC events and other, ``cascadelike'' events that predominantly consist of electromagnetic and hadronic showers.

The sample of events used in this work was collected in IceCube-DeepCore between April 2012 and May 2015. The event selection criteria only differ at the final selection level from those previously presented in~\cite{Aartsen:2017nmd} and~\cite{Aartsen:2019tjl} (sample ``$\mathcal{B}$'' therein). Starting from triggered events passing the DeepCore online filter~\cite{Collaboration:2011ym}, the selection applies coincidence and containment criteria, using the surrounding IceCube modules as active veto. This reduces both the rates of atmospheric $\mu^\pm$ and noise events by approximately eight orders of magnitude, leading to a sample with a purity of approximately \SI{95}{\percent} in atmospheric neutrinos and antineutrinos. 
To enhance sensitivity to NSI effects, we do not impose containment of the reconstructed stopping position of the event, and we keep all events whose reconstructed energy lies below \SI{100}{\giga\electronvolt} while adopting the same lower bound of \SI{5.6}{\giga\electronvolt} as previous analyses. All reconstructed zenith angles are accepted. 
We observe \num{47855} events in the sample, corresponding to an increase of $\sim \SI{15}{\percent}$ compared to sample ``$\mathcal{B}$.''
We employ identical methods as therein for obtaining the expected event distribution in all observable parameters from simulation, including detailed models of photon generation and propagation. The expected sample composition according to reaction channels at the best fit point within one of our NSI hypotheses is shown in Table~\ref{tab:event counts}, split up into the ``cascadelike'' and ``tracklike'' morphological categories that also constitute a binning dimension of the event histograms (cf. Sec.~\ref{subsec:statistical approach}).
For each event type, neutrinos dominate in the sample by a factor of about two to three times over antineutrinos, predominantly due to their larger cross sections~\cite{Honda:2015fha}. Furthermore, muon neutrinos contribute a significantly higher fraction of events (CC: $\sim \SI{60}{\percent}$) than do electron neutrinos (CC: $\sim \SI{23}{\percent}$) or, in particular, tau neutrinos (CC: $\sim \SI{3}{\percent}$).
Separation into ``tracklike'' and ``cascadelike'' events is based on the likelihood ratio obtained by reconstructing each event under two hypotheses: That of a cascade and track, expected for $\nu_\mu$ CC interactions, and that of a single cascade, expected for all other interactions.
The distribution of selected events in energy, direction and morphological category is shown in Fig.~\ref{fig:event histos} (Sec.~\ref{sec:results}).

\begin{table}[tbh]
\centering
\caption{\label{tab:event counts}Expected number of events for the best fit to data within the hypothesis of the generalized matter potential. The simulated events are broken down into all event types, including atmospheric $\mu^\pm$s. The numbers are split up into the ``cascadelike'' and ``tracklike'' event classification categories. The statistical uncertainties originate from limited simulation statistics.}
\begin{tabular}{lS[table-format=5.0]@{\,\( \pm \)\,}
    S[table-format=2.0]S[table-format=5.0]@{\,\( \pm \)\,}
    S[table-format=2.0]}
\toprule\toprule
Event type & \multicolumn{2}{c}{Cascadelike} & \multicolumn{2}{c}{Tracklike}\\
\midrule
$\nu_e$ CC & 5756 & 20 & 1799 & 11 \\
$\bar{\nu}_e$ CC & 2481 & 13 & 765 & 7\\
$\nu_\mu$ CC & 9811 & 27 & 9429 & 27 \\
$\bar{\nu}_\mu$ CC & 4328 & 18 & 4935 & 20 \\
$\nu_\tau$ CC & 835 & 7 & 317 & 4 \\
$\bar{\nu}_\tau$ CC & 374 & 5 & 144 & 3\\
$\nu_e$ NC & 465 & 6 & 141 & 3 \\
$\bar{\nu}_e$ NC & 135 & 3 & 43 & 2 \\
$\nu_\mu$ NC & 1731 & 11 & 569 & 7 \\
$\bar{\nu}_\mu$ NC & 584 & 7 & 193 & 4 \\
$\nu_\tau$ NC & 342 & 4 & 104 & 3 \\
$\bar{\nu}_\tau$ NC & 93 & 2 & 31 & 1 \\
Background ($\mu^\pm$) & 1187 & 37 & 1353 & 38 \\
\midrule
Total predicted & 28123 & 57 & 19823 & 53 \\
\midrule
Total observed & \multicolumn{2}{c}{28202} & \multicolumn{2}{c}{19653} \\
\bottomrule\bottomrule
\end{tabular}
\end{table}

\section{\label{sec:analysis}Analysis}

\subsection{\label{subsec:models}NSI hypotheses}
The IceCube DeepCore event sample introduced in Sec.~\ref{sec:DeepCore event selection} is interpreted assuming Schr{\"o}dinger-like evolution of three active neutrinos and six different matter Hamiltonians. Each represents a distinct NSI hypothesis, as summarized in Table~\ref{tab:NSI models}.

\begin{table}[tbh]
  \caption{\label{tab:NSI models}NSI hypotheses studied in this analysis of the IceCube DeepCore event sample detailed in Sec.~\ref{sec:DeepCore event selection}. While the first two hypotheses allow only for lepton flavor nonuniversality, the following three allow only for lepton flavor violation. The last one, based on the generalized matter potential parametrization of the matter Hamiltonian in Eq.~(\ref{eq:vacuum Hamiltonian like parameterisation}), places less restrictions on the NSI flavor structure.}
  \centering
    \begin{tabular}{lccc}
    \toprule
    \toprule
    Hypothesis & Parameters & Sampling grid\\
    \midrule
    $e$-$\mu$ NU & $\epsilon^\oplus_{ee} - \epsilon^\oplus_{\mu\mu}$ & $[-5, 5]$\\
    $\mu$-$\tau$ NU & $\epsilon^\oplus_{\tau\tau} - \epsilon^\oplus_{\mu\mu}$ & $[-0.10, 0.10]$\\
    $e$-$\mu$ FV & $\abs{\epsilon^\oplus_{e\mu}}$, $\delta_{e\mu}$ & $[0, 0.30] \times [\SI{0}{\degree}, \SI{360}{\degree}]$ \\
    $e$-$\tau$ FV & $\abs{\epsilon^\oplus_{e\tau}}$, $\delta_{e\tau}$ & $[0, 0.35] \times [\SI{0}{\degree}, \SI{360}{\degree}]$\\
    $\mu$-$\tau$ FV & $\abs{\epsilon^\oplus_{\mu\tau}}$, $\delta_{\mu\tau}$ & $[0, 0.07] \times [\SI{0}{\degree}, \SI{360}{\degree}]$\\
    GMP & $\epsilon_\oplus$, $\varphi_{12}$, $\varphi_{13}$ & $[-10, 10] \times [\SI{-90}{\degree}, \SI{90}{\degree}]^2 $\\
    \bottomrule\bottomrule
    \end{tabular}%
\end{table}%

The five phenomenological NSI parameters defined in Eqs.~(\ref{eq:real NU couplings}) and~(\ref{eq:complex FC couplings}) are assumed to be nonzero ``one-by-one'' with the four remaining parameters fixed to zero in each case. Here we follow the convention in the literature of constraining NSI coupling strengths by allowing only one to be nonzero at a time~\cite{Farzan:2017xzy}. This method is necessarily model dependent; the most generally applicable constraints result from accounting for the correlations between all couplings. These correlations can lead to partial cancellations and thereby to weakened constraints compared to those resulting from assuming one coupling at a time~\cite{Friedland:2004ah}. Nevertheless, we take this approach in the first part of this NSI search, not least because there are several theoretical NSI models that accommodate the possibility of the existence of only a single or a small number of sizable coupling strengths relevant to neutrino propagation~\cite{PhysRevD.79.011301,Ohlsson:2009vk,Heeck:2018nzc}. Unlike this work, none of the previous analyses using IceCube data~\cite{Esmaili:2013fva,Salvado:2016uqu,Aartsen:2017xtt,Demidov:2019okm} have performed measurements of $e$-$\mu$ nonuniversality or of complex couplings.

Testing the generalized matter potential in Eq.~(\ref{eq:vacuum Hamiltonian like parameterisation}) with three nonzero parameters $\epsilon_\oplus$, $\varphi_{12}$, and $\varphi_{13}$ has a reduced model dependence compared to the one-by-one fits. Moreover, probing all dimensions of this parameter space simultaneously is computationally feasible within our frequentist statistical framework due to the reduced dimensionality with respect to the parametrization in Eqs.~(\ref{eq:real NU couplings}) and~(\ref{eq:complex FC couplings}).

\subsection{\label{subsec:statistical approach}Statistical approach}
We perform a $\chi^2$ fit to a histogram of the observed events, binned in the reconstructed $\overset{\brabarb}{\nu}\vphantom{\nu}$ energy $E_\mathrm{reco}$, cosine-zenith $\cos(\vartheta_\mathrm{reco})$, and event type. For $E_\mathrm{reco}$, we employ eight bins covering the range from $10^{0.75} = \SI{5.6}{\giga\electronvolt}$ to $10^{1.75} = \SI{56.2}{\giga\electronvolt}$ that are uniformly spaced in $\log_{10}\left(E_\mathrm{reco}/\si{\giga\electronvolt}\right)$, extended by one bin reaching up to \SI{100}{\giga\electronvolt}. For $\cos(\vartheta_\mathrm{reco})$, we divide the range from \numrange{-1}{1} into eight uniformly spaced bins. The third histogram dimension is divided into the two flavor classification categories introduced in Sec.~\ref{sec:DeepCore event selection}, namely cascadelike and tracklike events. In total, there are $N_\mathrm{bins} = 9 \times 8 \times 2 = 144$ bins.

Each hypothesis from Sec.~\ref{subsec:models} is fit to the three-dimensional event histogram through the minimization of a modified Pearson's $\chi^2$ function defined as in~\cite{Aartsen:2017nmd,Aartsen:2019tjl} (see App~\ref{app:mod chi2} for more detail).
All fits are performed by finding the global $\chi^2$ minimum $\chi^2_\mathrm{min} \equiv \min \chi^2$ in the multidimensional space of NSI and nuisance parameters. 
The $d=1,2,3$-dimensional space defined by the respective NSI parameters\footnote{With $d=1$ for flavor-diagonal parameters, $d=2$ for flavor-violating complex parameters, and $d=3$ for the GMP parametrization.} is furthermore mapped-out using a dense grid of the same dimension, consisting of $N_g$ points $\{\bm{g}\}_{i=1, \dots, N_g} \in \mathcal{C}^d $. From the difference between the $\chi^2$ values resulting from minimizing with NSI parameter values fixed to the single grid points, $\{\chi^2_{\mathrm{min}}\left(\bm{g}_i\right)\}_{i=1, \dots, N_g}$, and the global $\chi^2_\mathrm{min}$, $\Delta \chi^2$ profiles are obtained.

The sampling grids for all six hypotheses are specified in Table~\ref{tab:NSI models}. In each case, the number of points $N_g$ is of the order of $10^{2d}$.

Due to the computational infeasibility\footnote{The total computational cost of this exceeds the available resources by approximately one order of magnitude, driven by the large number of hypotheses and complex minimization.} of a Feldman-Cousins approach~\cite{Feldman:1997qc}, we derive $d$-dimensional frequentist confidence regions by applying Wilks' theorem to a given $\Delta \chi^2$ profile, i.e., by assuming that it behaves as a $\chi^2$ distribution with $d$ degrees of freedom~\cite{algeri2019searching}. In the case $d=1$, these confidence regions correspond to confidence intervals on the sampled NSI parameters. When $d = 2$ or $d = 3$, we determine the confidence regions and intervals in all $d = 1$ and $d = 2$ parameter subsets from the projections of the original, higher-dimensional $\Delta \chi^2$ profile.

In order to prevent the lower-dimensional projections from getting biased due to the discrete nature of the samples in the NSI parameters to be optimized, the following routine is employed. For each point in the NSI parameters onto which the high-dimensional $\Delta \chi^2$ profile is to be projected, we search for local minima on the (one- or two-dimensional) grid spanning the space of NSI parameters that have to be optimized. Each local minimum that is detected is used as a seed for an additional local minimization process. The best fit among the set of minimization outcomes is recorded and employed in the projection.

\subsection{\label{subsec:nuisance}Nuisance parameters}
A total of 15 nuisance parameters are optimized in addition to each considered set of NSI parameters. This implies that the $\chi^2$ is a function of between 16 and 18 fit parameters, depending on the fit hypothesis. Table~\ref{tab:nuisance parameters} gives a list of the nuisance parameters found to be relevant throughout MC studies, grouped according to their origin. Each parameter is specified together with its prior constraints, where applicable, as well as its allowed fit range. Both the choice of prior and fit range include our understanding of the behavior of the respective parameter. In addition, the fit ranges are restricted to avoid unphysical parameter space.
\begin{table}[tbh]
  \centering
    \caption{
    \label{tab:nuisance parameters}Nuisance parameters employed by all NSI fits, as well as their associated Gaussian priors and fit ranges. For a given parameter with a prior, the range is specified as a number of standard deviations ($\sigma$) from the prior's nominal value. See text for the interpretation of all parameters.
    }
    \begin{tabular}{lcc}
    \toprule
    \toprule
    Parameter & Prior & Fit range\\
    \midrule
    \multicolumn{3}{c}{\textit{$\overset{\brabarb}{\nu}\vphantom{\nu}$ flux \& cross section:}}\\
    $(\nu_e + \bar{\nu}_e)/(\nu_\mu + \bar{\nu}_\mu)$ ratio & $1.00 \pm 0.05$ & $\pm 3\sigma$ \\
    $\nu/\bar{\nu}$ ratio ($\sigma$) & $0.0 \pm 1.0$ & $\pm 3\sigma$\\
    $\Delta\gamma_\nu$ (spectral index) & $0.0 \pm 0.1$ & $\pm 3\sigma$\\
    Effective livetime (years) & $\cdots$ & $[0, 3.8]$\\
    $M_A^\mathrm{CCQE}$ (quasielastic) (\si{\giga\electronvolt}) & $0.99^{+0.25}_{-0.15}$ & $\pm 3\sigma$\\
    $M_A^\mathrm{res}$ (resonance) (\si{\giga\electronvolt}) & $1.12 \pm 0.22$ & $\pm 3\sigma$\\
    NC normalization & $1.0 \pm 0.2$ & $\pm 3\sigma$ \\
    \multicolumn{3}{c}{\textit{Oscillation:}}\\
    $\theta_{23}$ (\si{\degree}) & $\cdots$ & $[30, 60]$\\
    $\Delta m^2_{32}$ ($\times \SI{e-3}{\electronvolt\squared}$) & $\cdots$  & $[0.93, 3.93]$\\
    \multicolumn{3}{c}{\textit{Detector:}}\\
    Optical efficiency, overall (\si{\percent}) & $100 \pm 10$ & $\pm 2\sigma$ \\
    Optical efficiency, lateral ($\sigma$) & $0.0 \pm 1.0$ & $^{+2.5\sigma}_{-2\sigma}$\\
    Optical efficiency, head-on (a.u.) & $\cdots$ & $[-5, 2]$\\
    Bulk ice, scattering (\si{\percent}) & $100 \pm 10$ & $\pm 1\sigma$\\
    Bulk ice, absorption (\si{\percent}) & $100 \pm 10$ & $\pm 1\sigma$\\
    \multicolumn{3}{c}{\textit{Atmospheric muons:}}\\
    Atmospheric muon fraction (\si{\percent}) & $\cdots$ & $[0, 35]$\\
    \bottomrule\bottomrule
    \end{tabular}
\end{table}
We account for seven uncertainties related to the intrinsic flux of atmospheric neutrinos and their detection cross sections, where the unconstrained effective livetime represents several uncertainties related to the overall normalization of the observed event count.

Of the six vacuum Hamiltonian parameters, we only let $\theta_{23}$ and $\Delta m^2_{32}$ vary, without imposing any prior constraints. The remaining parameters have small impact on the event sample under study and are fixed to $\theta_{12}=\SI{33.62}{\degree}$, $\theta_{13}=\SI{8.54}{\degree}$, $\Delta m^2_{21}=\SI{7.40e-5}{\electronvolt\squared}$~\cite{nufit3,nufit3.2url}, and $\delta_\mathrm{CP}=0$.

The detector related uncertainties include optical properties of the deep glacial ice and the photosensors' efficiency of detecting Cherenkov photons---both overall and depending on their angle of incidence.

The normalization of the atmospheric muon background distribution, given as a fraction of the total size of the event sample, is also included as an unconstrained nuisance parameter.

A more detailed interpretation of all nuisance parameters can be found in~\cite{Aartsen:2019tjl}. This also includes nuisance parameters that were found to be negligible, such as the upward-going vs. horizontal flux of electron neutrinos and local ice properties. 

\section{\label{sec:results}Results}
Table~\ref{tab:result summary} gives an overview of the outcomes of the six separate NSI fits discussed in Sec.~\ref{subsec:models}. The outcome of fitting SI is shown in addition in order to set the null hypothesis which is nested within all NSI hypotheses. All fits are performed within the parameter space of the normal ordering, i.e. $\Delta m^2_{32} > 0$. Depending on the NSI hypothesis under consideration in the respective fit, this choice does not \textit{a priori} result in any loss of generality of the derived NSI constraints. We return to the mass ordering question below in the context of each set of fit results.

All outcomes are characterized by a goodness of fit in the range of \SIrange{19}{22}{\percent}. The goodness of a given fit hypothesis is not determined from the $\Delta \chi^2$ value in Table~\ref{tab:result summary} but by comparing the observed value of $\chi^2_\mathrm{min}$ to the test-statistic distribution resulting from fitting the same hypothesis to a large number of statistically independent pseudoexperiments generated assuming SI. No nuisance parameter with an external constraint is found to experience a statistical pull from its best fit value beyond $1.1\sigma$, no matter which fit hypothesis is chosen (see Appendix~\ref{app:systematics} for more detail).

For computational reasons, the compatibility of the SI and NSI best fit hypothesis is tested using $\Delta \chi^2$ instead of pseudoexperiments. In all cases, the best fit SI hypothesis is statistically compatible with the best fit NSI hypothesis: the strongest disfavoring of the SI hypothesis is observed for the assumption of $e$-$\mu$ nonuniversality, at approximately $p = 0.3$. Our best fit values of the vacuum Hamiltonian parameters $\Delta m^2_{32}$ and $\theta_{23}$ under the SI hypothesis are compatible with the constraints found in the dedicated IceCube DeepCore analyses of Refs.~\cite{Aartsen:2017nmd,Aartsen:2019tjl}. In addition, the best fit values of $\Delta m^2_{32}$ and $\theta_{23}$ under the various NSI hypotheses are within \SI{2.5}{\percent} and \SI{4}{\percent}, respectively, of the values obtained assuming the SI hypothesis.
\begin{table}[tb]
  \centering
  \caption{\label{tab:result summary}Summary of fit outcomes for the NSI hypotheses considered in Table~\ref{tab:NSI models}, together with the best fit values of all NSI parameters, the $\Delta \chi^2$ values of the respective global $\chi^2_\mathrm{min}$ with respect to the SI hypothesis as well as the corresponding p-values. Since the matter potential has no free parameters for the SI case, we show the best fit values of the two considered vacuum Hamiltonian parameters $\Delta m^2_{32}$ and $\theta_{23}$ instead.}
    \begin{tabular}{lccc}
    \toprule\toprule
    Hypothesis & Best fit values & $\Delta \chi^2_\mathrm{SI}$ & p\\
    \midrule
    SI & $\Delta m^2_{32} = \SI{0.00237}{\electronvolt\squared}$, $\theta_{23} = \SI{46.4}{\degree}$ & \num{0} & \num{0}\\
    $e$-$\mu$ NU & $\epsilon^\oplus_{ee} - \epsilon^\oplus_{\mu\mu} = \num{-0.59}$ & \num{1.3} &\num{0.3}\\
    $\mu$-$\tau$ NU & $\epsilon^\oplus_{\tau\tau} - \epsilon^\oplus_{\mu\mu} = \num{0.0016}$ & \num{0.0} & \num{0.9}\\
    $e$-$\mu$ FV & $\abs{\epsilon^\oplus_{e\mu}} = \num{0.072}$, $\delta_{e\mu} = \SI{343.7}{\degree}$ & \num{1.2} & \num{0.3}\\
    $e$-$\tau$ FV & $\abs{\epsilon^\oplus_{e\tau}} = \num{0.060}$, $\delta_{e\tau} = \SI{35.5}{\degree}$ & \num{0.5}& \num{0.5}\\
    $\mu$-$\tau$ FV & $\abs{\epsilon^\oplus_{\mu\tau}} = \num{0.0030}$, $\delta_{\mu\tau} = \SI{175.0}{\degree}$ & \num{0.1} & \num{0.8} \\
    GMP & $\epsilon_\oplus = \num{0.40}$, $\varphi_{12} = \SI{2.3}{\degree}$, $\varphi_{13} = \SI{-4.7}{\degree}$ & \num{2.2} & \num{0.1}\\
    \bottomrule\bottomrule
    \end{tabular}%
\end{table}

\begin{figure*}[!tbh]
\centering
\includegraphics[width=\linewidth]{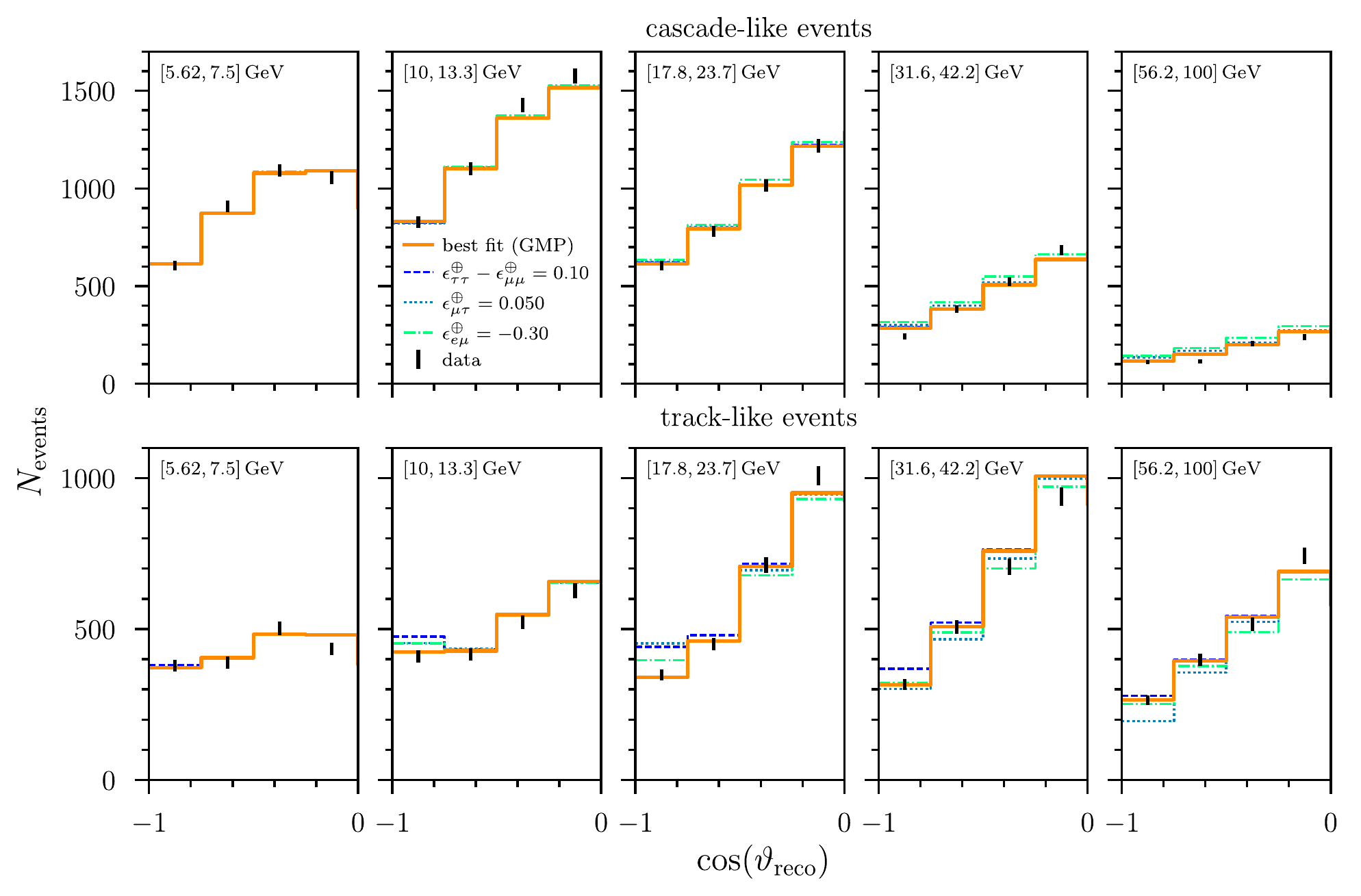}
\caption{\label{fig:event histos}Histograms of observed cascadelike events (top row) and tracklike events (bottom row) as a function of $\cos(\vartheta_\mathrm{reco})$ for different slices in $E_\mathrm{reco}$ (indicated at the top of each panel), together with the MC expectation under the generalized matter potential fit outcome, labeled as ``best fit (GMP).'' For display purposes, the eight lowest reconstructed energy bins have been merged into four, and only the upgoing region $\cos(\vartheta_\mathrm{reco}) \leq 0$ is shown, where the largest NSI effects are expected. Also shown are the expected event distributions for one particular $\mu$-$\tau$ nonuniversality realization ($\epsilon^\oplus_{\tau\tau} - \epsilon^\oplus_{\mu\mu} = 0.10$), one $\mu$-$\tau$ flavor violation realization ($\epsilon^\oplus_{\mu\tau} = 0.050$), and one $e$-$\mu$ flavor-violation realization ($\epsilon^\oplus_{e\mu} = -0.30$). In each of these three example NSI scenarios, all nuisance parameters are set to their respective global best fit values within the corresponding NSI parameter space.}
\end{figure*}

In Fig.~\ref{fig:event histos}, we show $E_\mathrm{reco}$ slices of the observed event counts as a function of $\cos(\vartheta_\mathrm{reco}) \leq 0$ for the two event classes (rows). In the figure, we have condensed the eight lowest energy bins into four slices, each of which covers two energy bins of the original binning used in the analysis. We also show the best fit of the generalized matter potential hypothesis, as well as three signal hypotheses with nonzero $\epsilon^\oplus_{\tau\tau}-\epsilon^\oplus_{\mu\mu}$, $\epsilon^\oplus_{\mu\tau}$, or $\epsilon^\oplus_{e\mu}$. For the latter three, all nuisance parameters are set to the values obtained by the respective global best fit, with the NSI coupling strengths given in Table~\ref{tab:result summary} (see Appendix~\ref{app:systematics} for the detailed nuisance parameter values). Thus, the induced event count differences follow solely from choosing different NSI parameter values compared to those that fit the data best; the event distributions at all three best fit points would be barely distinguishable by eye from the fit of the generalized matter potential in Fig.~\ref{fig:event histos}. 
These differences in count are what is observable of the imprints of NSI on the oscillation probability after superimposing the expected event distributions in $E_\mathrm{reco}$, $\cos(\vartheta_\mathrm{reco})$, and event classification of both the atmospheric $\mu^\pm$ background and the effective-area weighted oscillated fluxes of neutrinos and antineutrinos of all flavors; see the thirteen sample components in Table~\ref{tab:event counts}.

\subsection{\label{subsec:one-by-one results}One-by-one fits}
\subsubsection{\label{subsubsec:non-universal results}Flavor-nonuniversal NSI}
\begin{figure*}[bth]
	\begin{minipage}[c]{0.5\textwidth}
		\includegraphics[width=\linewidth]{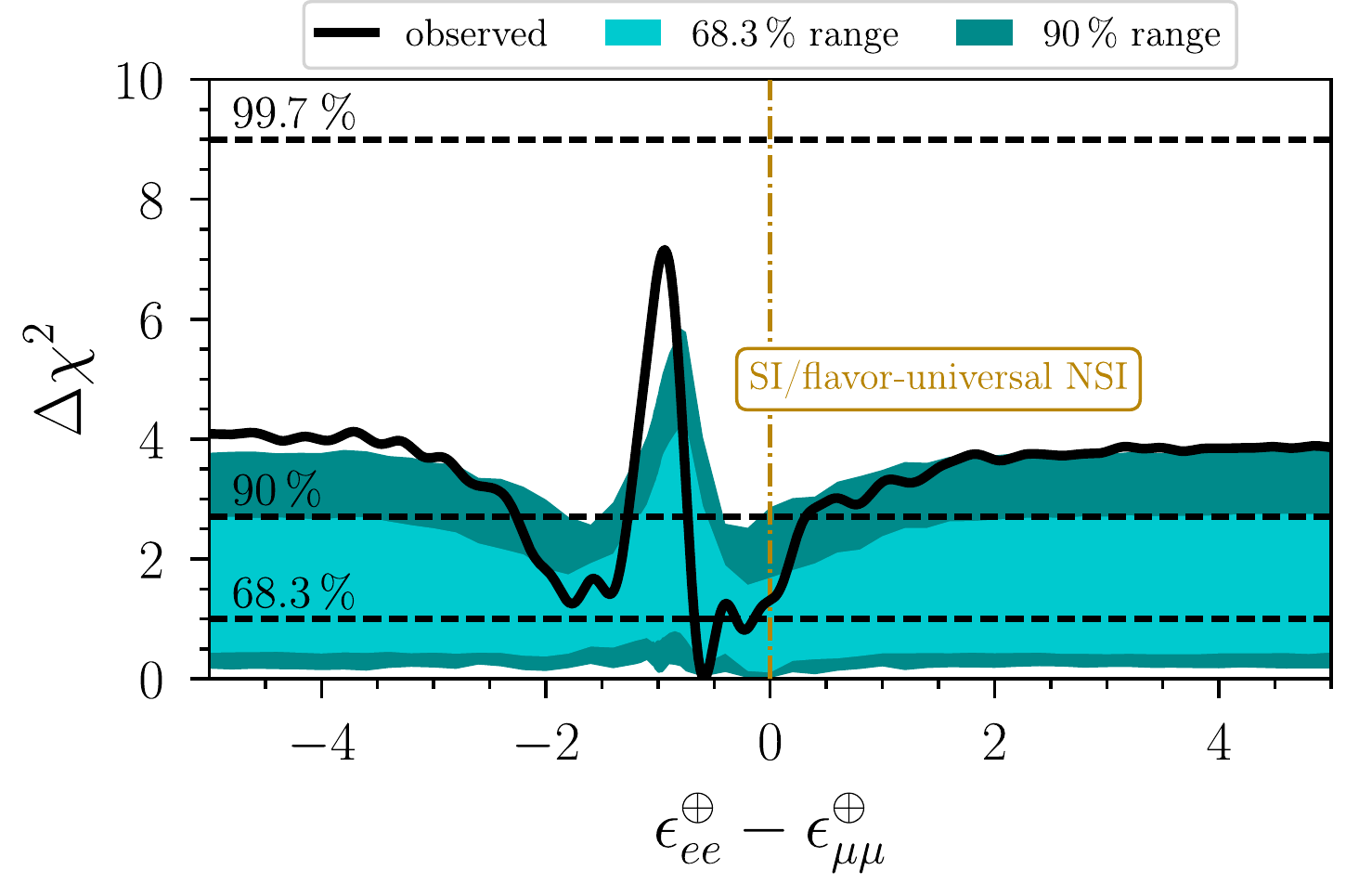}
	\end{minipage}\hfill
	\begin{minipage}[c]{0.5\textwidth}
		\includegraphics[width=\linewidth]{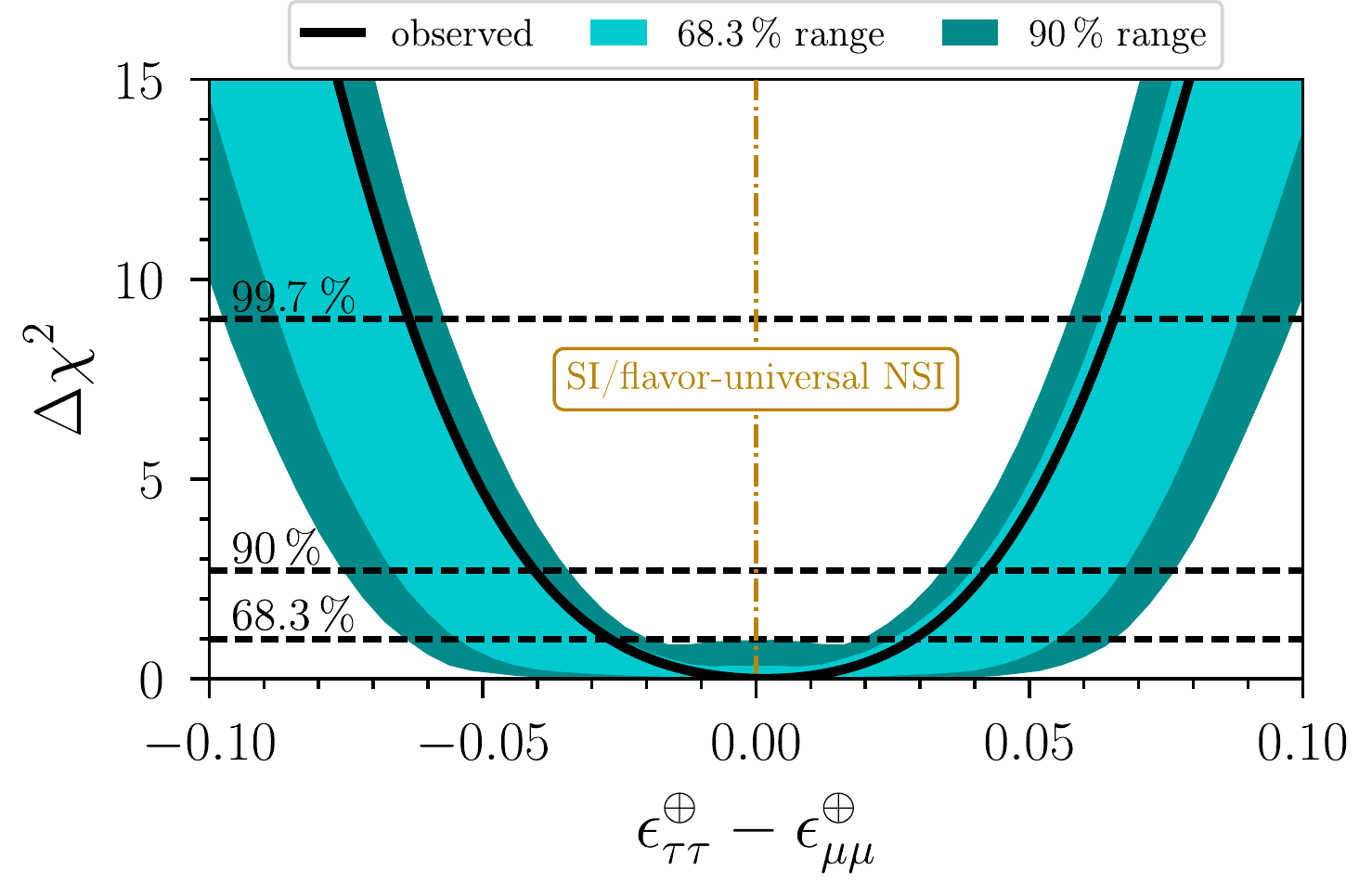}
	\end{minipage}
	\caption{Observed $\Delta \chi^2$ profiles as a function of the effective NSI flavor nonuniversality parameters $\epsilon^\oplus_{ee}-\epsilon^\oplus_{\mu\mu}$ (left) and $\epsilon^\oplus_{\tau\tau}-\epsilon^\oplus_{\mu\mu}$ (right), together with the central \SI{68.3}{\percent} and \SI{90}{\percent} confidence intervals of the experimental sensitivity shown as shaded bands. See text for details.}
	\label{fig:deltachi2 NU NSI}
\end{figure*}
Figure~\ref{fig:deltachi2 NU NSI} shows observed $\Delta \chi^2$ profiles as a function of the two differences of the flavor-diagonal NSI coupling strengths, namely $\epsilon^\oplus_{ee}-\epsilon^\oplus_{\mu\mu}$ and $\epsilon^\oplus_{\tau\tau}-\epsilon^\oplus_{\mu\mu}$. In each case, all other coupling strengths are fixed to zero.

The shaded bands give the experimental sensitivity by showing the symmetrical central \SI{68.3}{\percent} and \SI{90}{\percent} confidence intervals of the $\Delta \chi^2$ distributions obtained in fits to pseudoexperiments for the generation of which SI were assumed.

Horizontal dashed lines denote the 68.3th, 90th, and 99.7th percentiles of a $\chi^2$ distribution with one degree of freedom. 

Vertical dash-dotted lines mark the values of the two parameters that leave flavor transitions unchanged with respect to SI. Since neutrino oscillation experiments are not sensitive to the overall scale of the flavor-diagonal NSI coupling strengths, these lines represent both the SI hypothesis and the hypothesis of flavor-universal NSI.

\paragraph{$e$-$\mu$ nonuniversality}
The left panel of Fig.~\ref{fig:deltachi2 NU NSI} reveals that no constraints beyond $\Delta \chi^2 \approx 7.2$, corresponding to a confidence level (CL) of approximately \SI{99}{\percent}, can be placed on the $e$-$\mu$ nonuniversality parameter \mbox{$\epsilon^{\oplus}_{ee} - \epsilon^\oplus_{\mu\mu}$}. However, values outside of the union of intervals \mbox{$\left[-2.26, -1.27\right] \cup \left[-0.74, 0.32\right]$} are excluded at \SI{90}{\percent} CL. 

The vanishing impact from $\theta_{12}$ and $\delta_{CP}$ causes the sign of $1 + \epsilon^\oplus_{ee} - \epsilon^\oplus_{\mu\mu}$ to be fully degenerate with the mass ordering~\cite{Coloma:2016gei}. This applies similarly to $\mu$-$\tau$ nonuniversality and $|\epsilon^\oplus_{\mu\tau}|$ flavor-violation. We therefore do not need to test solutions within the inverted ordering explicitly. When interpreted in terms of standard matter effects, the $\Delta \chi^2$ profile asymmetry about $\epsilon^\oplus_{ee} - \epsilon^\oplus_{\mu\mu} = -1$ suggests that the data slightly favors the normal ordering, corresponding to the point $\epsilon^\oplus_{ee} - \epsilon^\oplus_{\mu\mu} = 0$, over the inverted ordering, corresponding to the point $\epsilon^\oplus_{ee} - \epsilon^\oplus_{\mu\mu} = -2$, at the level of $\Delta \chi^2 \approx 0.5$\footnote{See~\cite{Aartsen:2019eht} for a statistically rigorous study of the neutrino mass ordering in the absence of NSI with two related IceCube DeepCore event samples.}. A more detailed discussion of the profile characteristics and their causes can be found in Appendix~\ref{app:detailed eps ee flat sensitivity}.

\paragraph{$\mu$-$\tau$ nonuniversality}
From the right panel of Fig.~\ref{fig:deltachi2 NU NSI}, we find that the observed event sample is fully compatible with NSI that are $\mu$-$\tau$ flavor universal, that is, $\epsilon^\oplus_{\tau\tau}-\epsilon^\oplus_{\mu\mu}=0$. In contrast to the observed $\Delta \chi^2$ profile under the hypothesis of $e$-$\mu$ nonuniversality, here the test statistic keeps increasing for large values of $\abs{\epsilon^\oplus_{\tau\tau}-\epsilon^\oplus_{\mu\mu}}$, which allows for stringent constraints with values of $\epsilon^\oplus_{\tau\tau}-\epsilon^\oplus_{\mu\mu}$ outside the interval $[-0.041, 0.042]$ excluded by the data at \SI{90}{\percent} CL. The sensitivity to this type of NSI stems almost exclusively from its impact on the $\nu_\mu$ and $\bar{\nu}_\mu$ survival probabilities, cf. Sec.~\ref{subsec:NSI in oscillation probs}. We find that the summation over neutrinos and antineutrinos in general does not lead to significant cancellations of the respective NSI signatures.

\subsubsection{\label{subsubsec:flavour-changing results}Flavor-violating NSI}
\begin{figure*}[tbh]
	\begin{minipage}[c]{0.33\textwidth}
		\includegraphics[width=\linewidth]{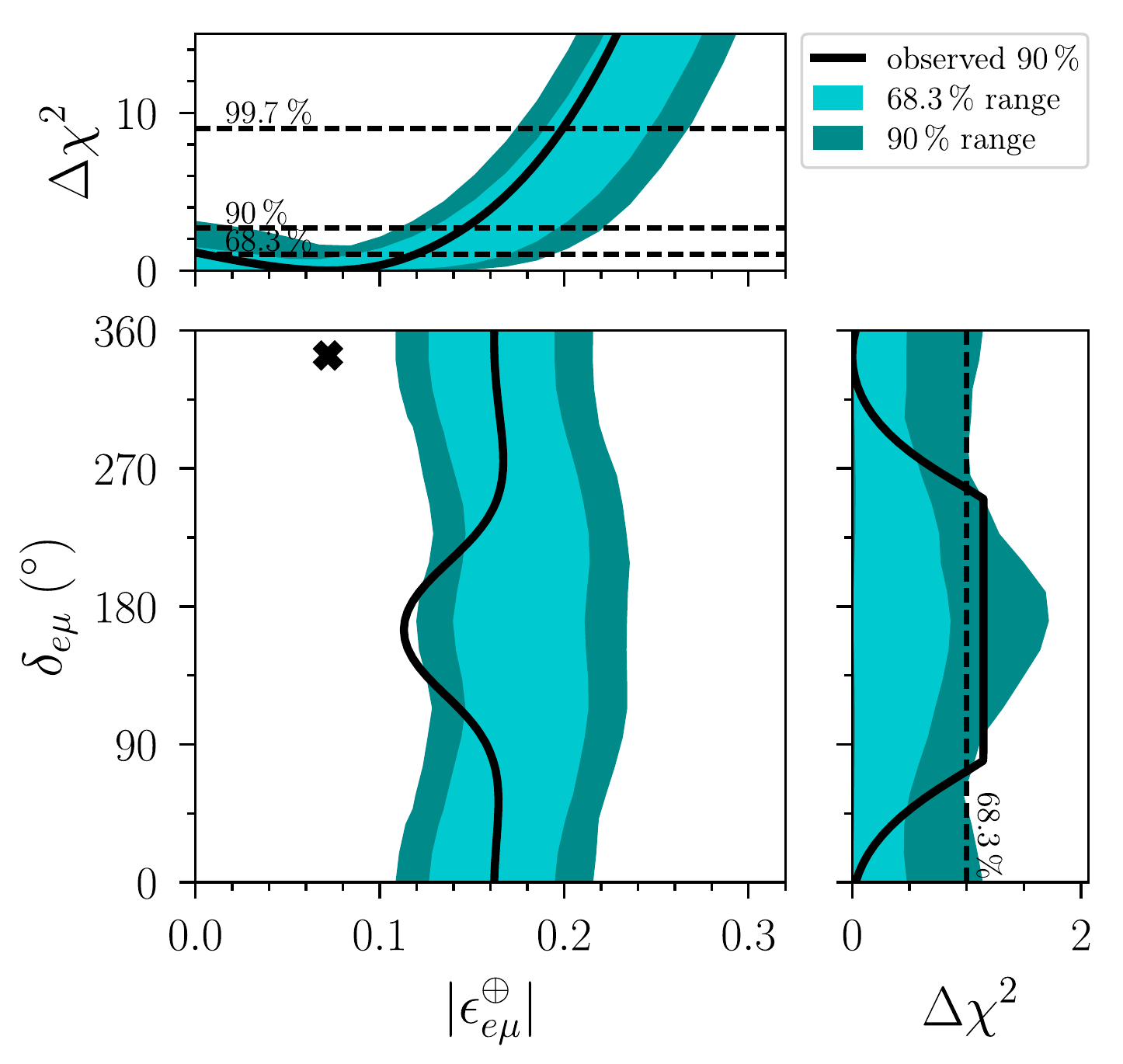}
	\end{minipage}\hfill
	\begin{minipage}[c]{0.33\textwidth}
		\includegraphics[width=\linewidth]{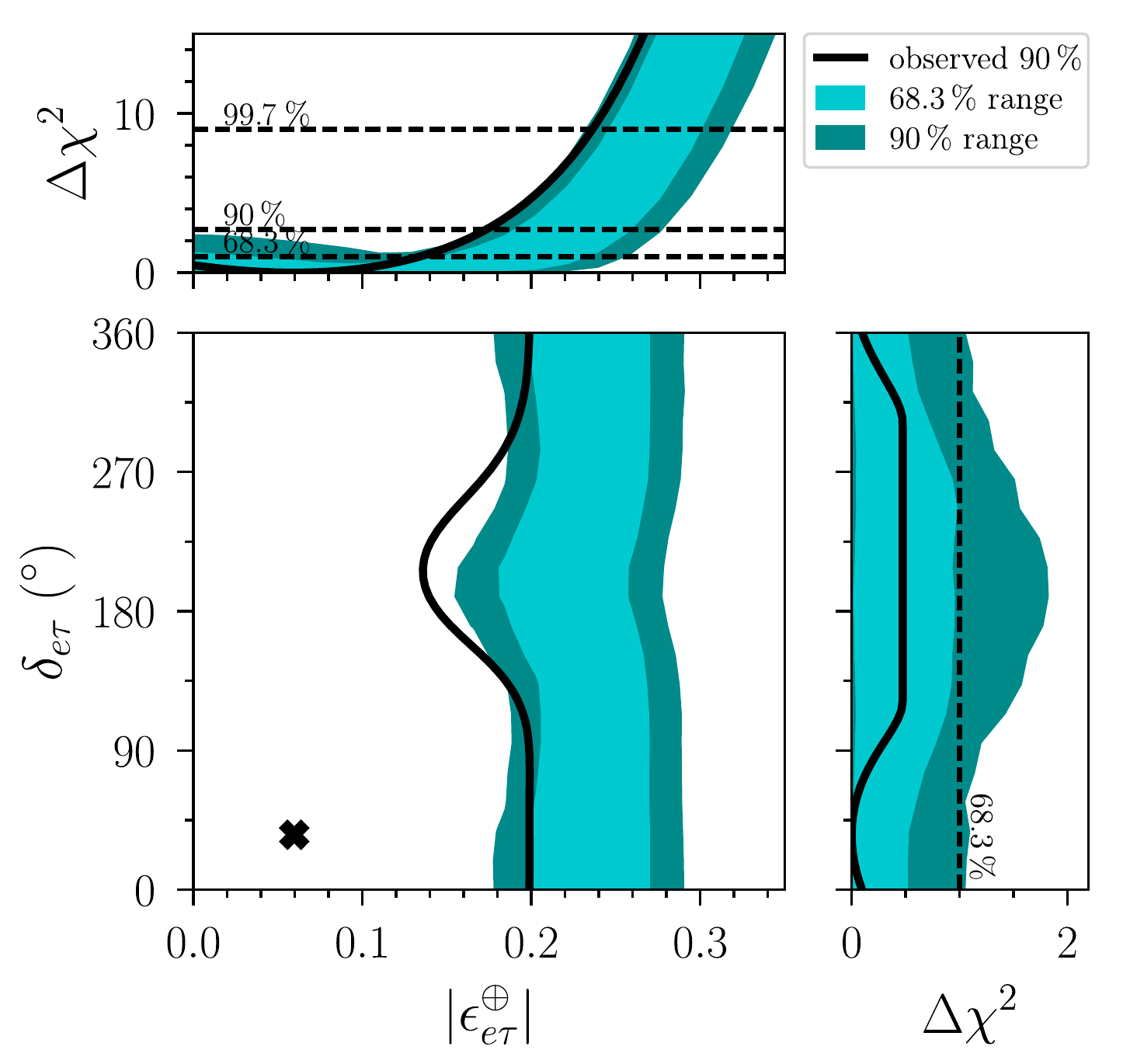}
	\end{minipage}
	\begin{minipage}[c]{0.33\textwidth}
		\includegraphics[width=\linewidth]{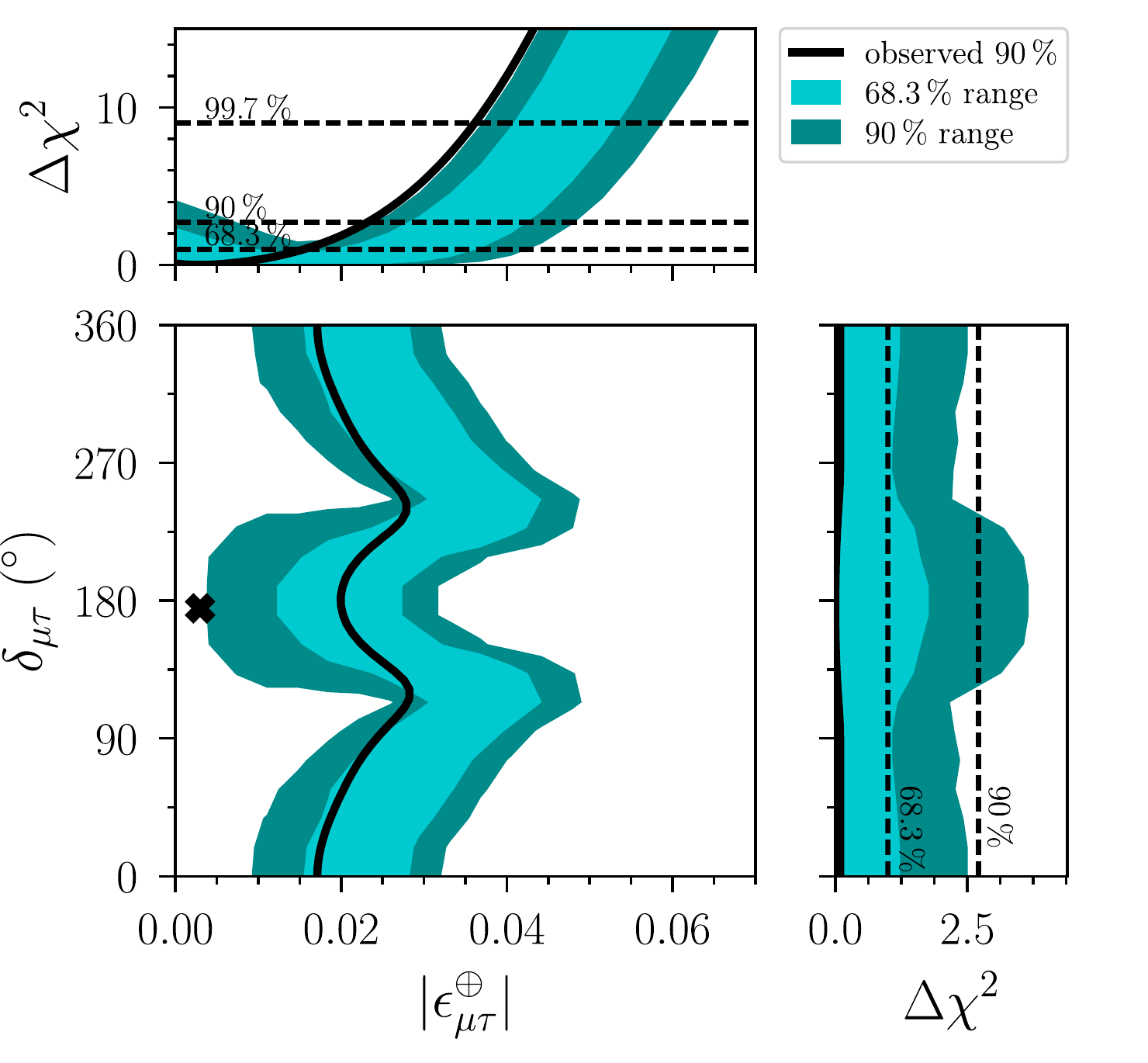}
	\end{minipage}
	\caption{Observed \SI{90}{\percent} confidence regions in the magnitudes $|\epsilon^\oplus_{\alpha\beta}|$ and phases $\delta_{\alpha\beta}$ of the effective flavor-violating NSI coupling strengths $\epsilon^\oplus_{e\mu}$ (left), $\epsilon^\oplus_{e\tau}$ (middle), and $\epsilon^\oplus_{\mu\tau}$ (right), together with each parameter's (projected) one-dimensional $\Delta \chi^2$ profile. The best fit point for each pair of parameters is indicated by a cross. The central \SI{68.3}{\percent} and \SI{90}{\percent} confidence regions and intervals of the experimental sensitivity are shown as shaded bands. See text for details.}
	\label{fig:deltachi2 FC NSI}
\end{figure*}
The central panel of the each of the three plots in Fig.~\ref{fig:deltachi2 FC NSI} shows the observed \SI{90}{\percent} CL contour ($\Delta \chi^2 \approx 4.61$) in the NSI magnitude and complex phase from the fit of a given flavor-violating NSI coupling strength. The projection of the two-dimensional $\Delta \chi^2$ profile onto the magnitude of the coupling strength is depicted on top, and that onto the complex phase on the right. Lines and shaded bands have the same meanings as in Fig.~\ref{fig:deltachi2 NU NSI}. Note that the SI case is located at the origin. The appropriate entry for $\Delta \chi^2_\mathrm{SI}$ in Table~\ref{tab:result summary} provides the maximal projected $\Delta \chi^2$ at which any value of the complex phase is disfavored, due to the projection method and the vanishing amplitude rendering the complex phase unphysical.

\paragraph{$e$-$\mu$ flavor violation}
From Fig.~\ref{fig:deltachi2 FC NSI} (left), the magnitude of $e$-$\mu$ flavor-violating NSI is compatible with zero (or SI) at a significance level of approximately $p = 0.3$ and an upper bound of $\abs{\epsilon^\oplus_{e\mu}} \leq 0.146$ (\SI{90}{\percent} CL) is obtained when the full range $\SI{0}{\degree} \leq \delta_{e\mu} \leq \SI{360}{\degree}$ is considered. A stronger constraint on the magnitude follows when $\delta_{e\mu}$ is only allowed to take more disfavored values of \SI{160}{\degree}$\pm$\SI{90}{\degree}. In this limited range of $\delta_{e\mu}$, the NSI magnitude that best fits the data is zero. This explains the ``plateau'' in the projection onto $\delta_{e\mu}$ with $\Delta \chi^2 = \Delta \chi^2_\mathrm{SI} \approx 1.1$ (compare Table~\ref{tab:result summary}). A somewhat stronger exclusion of real negative values of $\epsilon^\oplus_{e\mu}$ ($\delta_{e\mu} \approx \SI{180}{\degree}$) with respect to the expectation from pseudoexperiments for SI is observed.

Within the inverted ordering parameter space, the \SI{90}{\percent} CL exclusion contour shifts to larger values of $\abs{\epsilon^\oplus_{e\mu}}$ by approximately \SI{10}{\percent}; no change is observed for the one-dimensional $\abs{\epsilon^\oplus_{e\mu}}$ interval allowed at \SI{90}{\percent} when the inverted ordering is adopted. Furthermore, the one-dimensional allowed intervals (at any CL) for both parameters obtained under the assumption of $\Delta m^2_{32} > 0$ also apply to the scenario in which the mass ordering is considered as a nuisance parameter. For the observed $e$-$\tau$ flavor-violation constraints switching from the normal ordering to the inverted ordering parameter space similarly has negligible impact.

\paragraph{$e$-$\tau$ flavor violation}
Compared to $e$-$\mu$ flavor violation, we find both qualitatively and quantitatively similar bounds on $e$-$\tau$ flavor violation; see Fig.~\ref{fig:deltachi2 FC NSI} (middle). The \SI{90}{\percent} CL upper bound on the NSI magnitude from optimizing over $\SI{0}{\degree} \leq \delta_{e\tau} \leq \SI{360}{\degree}$ is slightly larger, $\abs{\epsilon^\oplus_{e\tau}} \leq 0.173$, and the best fit is well compatible with the SI hypothesis. For $\delta_{e\tau}$ values in an approximately \SI{90}{\degree} range around \SI{200}{\degree}, a somewhat more constraining bound results for the magnitude. The NSI hypotheses corresponding to this limited range of $\delta_{e\tau}$ best fit the data when $\abs{\epsilon^\oplus_{e\tau}} = 0$, leading to a plateau in the projection onto $\delta_{e\tau}$ with $\Delta \chi^2 = \Delta \chi^2_\mathrm{SI} \approx 0.5$.

\paragraph{$\mu$-$\tau$ flavor violation}
The right side of Fig.~\ref{fig:deltachi2 FC NSI} suggests that the selected event sample has significantly better sensitivity to $\mu$-$\tau$ flavor violation than to flavor violation in the electron sector considered in Sec.~\ref{subsec:NSI in oscillation probs}.
When the full $\delta_{\mu\tau}$ range is allowed, $\abs{\epsilon^\oplus_{\mu\tau}} \leq 0.0232$ at \SI{90}{\percent} CL. We find the strongest bounds on the NSI magnitude for real NSI, i.e., for $\delta_{\mu\tau} = \SI{0}{\degree}$ and $\delta_{\mu\tau} = \SI{180}{\degree}$. The data sample shows almost vanishing sensitivity to $\delta_{\mu\tau}$. This observation is in agreement with Table~\ref{tab:result summary}, which gives $\Delta \chi^2_\mathrm{SI} \approx 0.1$ as the maximum of the $\Delta \chi^2$ projection onto $\delta_{\mu\tau}$. Hypotheses with $\delta_{\mu\tau} \approx \SI{125}{\degree}$ and $\delta_{\mu\tau} \approx \SI{235}{\degree}$ result in the weakest constraints on the magnitude, not those with $\delta_{\mu\tau} = \SI{90}{\degree}$ or $\delta_{\mu\tau} = \SI{270}{\degree}$, for which the contribution of the magnitude $\abs{\epsilon^\oplus_{\mu\tau}}$ to the oscillation probabilities in the $\mu$-$\tau$ sector is reduced, see Sec.~\ref{subsec:NSI in oscillation probs}. Pseudoexperiments suggest that such a deviation  is characteristic of considering the joint $\nu_\mu$ CC and $\bar{\nu}_\mu$ CC event distribution. Indeed, the strength of cancellations between the effects of $\abs{\epsilon^\oplus_{\mu\tau}}$ on neutrino and antineutrino channels in the medium- to high-energy regime depends on the value of $\delta_{\mu\tau}$.

\subsubsection{\label{subsec:one-by-one results summary}Summary and experiment comparison}

{\renewcommand{\arraystretch}{1.2}
\begin{table}[t]
  \centering
  \caption{\label{tab:constraints summary}Summary of \SI{90}{\percent} CL constraints on NSI nonuniversality and flavor-violation parameters obtained by the one-by-one fits in this study, as well as on the parameters of the generalized matter potential, whose fit is discussed in Sec.~\ref{subsec:generalised matter potential}. $\Delta m^2_{32} > 0$ is assumed everywhere, but does not introduce a loss of generality (see text for details).}
    \begin{tabular}{lcc}
    \toprule\toprule
    Hypothesis & Parameter & Allowed interval (\SI{90}{\percent} CL)\\
    \midrule
    $e$-$\mu$ NU & $\epsilon^\oplus_{ee} - \epsilon^\oplus_{\tau\tau}$ & $\left[-2.26, -1.27\right] \cup \left[-0.74, 0.32\right]$\\
    $\mu$-$\tau$ NU & $\epsilon^\oplus_{\tau\tau} - \epsilon^\oplus_{\mu\mu}$ & $\left[-0.041, 0.042\right]$\\
    \multirow{2}{*}{$e$-$\mu$ FV} & $\abs{\epsilon^\oplus_{e\mu}}$ & $\leq 0.146$ \\
    & $\delta_{e\mu}$ & $\left[\SI{0}{\degree},\SI{360}{\degree}\right]$\\
    \multirow{2}{*}{$e$-$\tau$ FV} & $\abs{\epsilon^\oplus_{e\tau}}$ & $\leq 0.173$ \\
    & $\delta_{e\tau}$ & $\left[\SI{0}{\degree},\SI{360}{\degree}\right]$\\
    \multirow{2}{*}{$\mu$-$\tau$ FV} & $\abs{\epsilon^\oplus_{\mu\tau}}$ & $\leq 0.0232$ \\
    & $\delta_{\mu\tau}$ & $\left[\SI{0}{\degree},\SI{360}{\degree}\right]$\\
    \midrule
    \multirow{3}{*}{GMP} & $\epsilon_\oplus$ & $\left[-1.2, -0.3\right] \cup \left[0.2, 1.4\right]$ \\
    & $\varphi_{12}$ & $\left[\SI{-9}{\degree}, \SI{8}{\degree}\right]$\\
    & $\varphi_{13}$ & $\left[\SI{-14}{\degree}, \SI{9}{\degree}\right]$\\
    \bottomrule\bottomrule
    \end{tabular}
\end{table}
}

Table~\ref{tab:constraints summary} compiles a summary of the constraints (at \SI{90}{\percent} CL) placed by this analysis on the NSI flavor nonuniversality and flavor-violation parameters. Both SI and flavor-universal NSI (in the case of flavor-diagonal couplings) are compatible with each best fit NSI hypothesis. None of the complex phases are constrained at \SI{90}{\percent} CL.

\begin{figure}[t]
\centering
\includegraphics[width=1.0\linewidth]{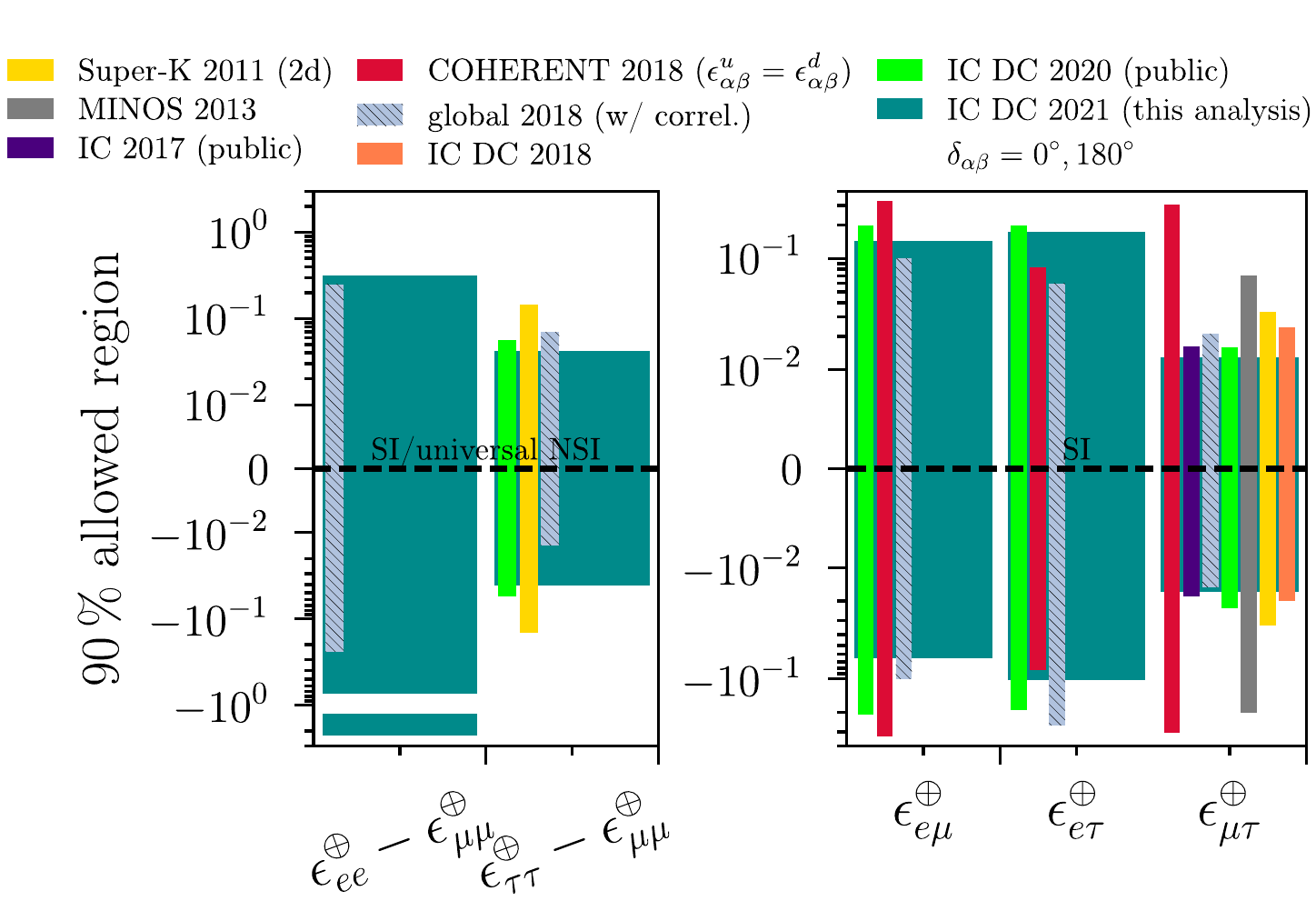}
\caption{\label{fig:constraints summary}Summary of the one-by-one constraints at \SI{90}{\percent} CL on real NSI nonuniversality and flavor-violation parameters obtained in this study (labeled as ``IC DC 2021'') compared to previous limits~\cite{PhysRevD.84.113008,PhysRevD.88.072011,Salvado:2016uqu,Denton:2018xmq,Esteban:2018ppq,Aartsen:2017xtt,Demidov:2019okm}. Constraints on the magnitudes of complex NSI parameters are given for the respective phase restricted to $\delta_{\alpha\beta} = $\SI{0}{\degree}, \SI{180}{\degree}. See text for details.}
\end{figure}

For comparison with existing measurements, in Fig.~\ref{fig:constraints summary} we restrict the flavor-violating coupling strengths to the real plane, defined by $\delta_{\alpha\beta} = \SI{0}{\degree},\SI{180}{\degree}$, and show the \SI{90}{\percent} CL intervals for the real-valued signed coupling strengths $\epsilon^\oplus_{\alpha\beta}$. The lower limits ($\delta_{\alpha\beta} = \SI{180}{\degree}$) on $\epsilon^\oplus_{e\mu}$ and $\epsilon^\oplus_{e\tau}$ are stronger than their upper limits ($\delta_{\alpha\beta} = \SI{0}{\degree}$). The latter reproduce the constraints on the NSI magnitudes $\abs{\epsilon^\oplus_{e\mu}}$ and $\abs{\epsilon^\oplus_{e\tau}}$ that are found under the hypotheses of complex coupling strengths in Table~\ref{tab:constraints summary}. In the case of $\epsilon^\oplus_{\mu\tau}$, the upper limit is slightly stronger than the lower limit, $-0.0165 \leq \epsilon^\oplus_{\mu\tau} \leq 0.0130$. This range is fully compatible with, and smaller than, that reported by our previous study, $-0.020 \leq \epsilon^\oplus_{\mu\tau} \leq 0.024$~\cite{Aartsen:2017xtt}.\footnote{After translating from NSI with down quarks to the effective coupling strengths in Eq.~(\ref{eq:constant Earth effective couplings protons neutrons}).} Neither of the magnitudes of the upper or the lower limit reproduces the limit on $\abs{\epsilon^\oplus_{\mu\tau}}$ in Table~\ref{tab:constraints summary} because the sensitivity of the event sample to $\abs{\epsilon^\oplus_{\mu\tau}}$ is weakest for a complex coupling strength (cf. Fig.~\ref{fig:deltachi2 FC NSI}).

Data from a number of other neutrino experiments has been used to set limits on the NSI coupling strengths $\epsilon^{uV,dV}_{\alpha\beta}$, which we have rescaled for consistency with the definition of the effective coupling strengths for Earth matter, Eq.~(\ref{eq:constant Earth effective couplings protons neutrons}). Figure~\ref{fig:constraints summary} contains results reporting one-dimensional \SI{90}{\percent} CL intervals, almost all of which are based on one-by-one fits similar to those discussed in Sec.~\ref{subsec:one-by-one results}. Among these are limits on $\epsilon^\oplus_{\tau\tau} - \epsilon^\oplus_{\mu\mu}$ and $\epsilon^\oplus_{\mu\tau}$ (with correlations) obtained from atmospheric neutrino data collected by Super-Kamiokande~\cite{PhysRevD.84.113008}, as well as limits on $\epsilon^\oplus_{\mu\tau}$ from long-baseline accelerator $\overset{\brabarb}{\nu}\vphantom{\nu}_\mu$ disappearance data from MINOS~\cite{PhysRevD.88.072011} and high-energy atmospheric $\overset{\brabarb}{\nu}\vphantom{\nu}_\mu$ disappearance data from IceCube~\cite{Salvado:2016uqu} (labeled ``IC 2017''), respectively. Furthermore, we show the limits on flavor-violating coupling strengths reported by an analysis of the published timing (or flavor) data from coherent elastic neutrino-nucleus scattering (CE$\nu$NS) at COHERENT~\cite{Denton:2018xmq}. Here, the assumed underlying NSI model based on the exchange of a $Z^\prime$ mediator with $M_{Z^\prime} \sim \order{\SI{10}{\mega\electronvolt}}$ dictates $\epsilon^u_{\alpha\beta} = \epsilon^d_{\alpha\beta}$, so that no cancellations between NSI with different quark flavors occur (see~\cite{PhysRevD.101.035039} for a comprehensive analysis).

While CE$\nu$NS only yields constraints that are valid for a new physics energy scale above $\order{\SI{10}{\mega\electronvolt}}$, it is sensitive to the individual flavor-diagonal coupling strengths $\epsilon^{uV,dV}_{ee}$ and  $\epsilon^{uV,dV}_{\mu\mu}$ (not depicted in Fig.~\ref{fig:constraints summary})---in contrast to neutrino oscillation experiments.
Similarly, our results are not directly comparable to NSI limits set in collider experiments as these commonly depend strongly on the underlying model and new physics energy scale~\cite{BABU2021136131}.

Figure~\ref{fig:constraints summary} additionally allows gauging the impact of the increased event statistics and the inclusion of higher-energy events in our sample compared to a study with public IceCube DeepCore data in~\cite{Demidov:2019okm}, labeled ``IC DC 2020 (public).'' The widths of the \SI{90}{\percent} CL intervals are smaller by between $\sim \SI{25}{\percent}$ (for $\epsilon^\oplus_{\tau\tau} - \epsilon^\oplus_{\mu\mu}$, $\epsilon^\oplus_{e\tau}$, $\epsilon^\oplus_{\mu\tau}$) and $\sim \SI{50}{\percent}$ ($\epsilon^\oplus_{e\mu}$).

We also display the limits derived in a combined analysis of global neutrino oscillation datasets with negligible sensitivity to $CP$-violating effects~\cite{Esteban:2018ppq} (``global 2018''). The global analysis only assumes the coupling strengths $\epsilon^{uV,dV}_{\alpha\beta}$ to be nonzero; their flavor structure is taken to be independent of the quark type. The fact that correlations between NSI couplings with different flavor indices are fully taken into account explains why these constraints are no more stringent than those found in this study.

\subsection{\label{subsec:generalised matter potential}Generalized matter potential}
\begin{figure*}[tb]
	\begin{minipage}[c]{0.5\textwidth}
		\includegraphics[width=.9\linewidth]{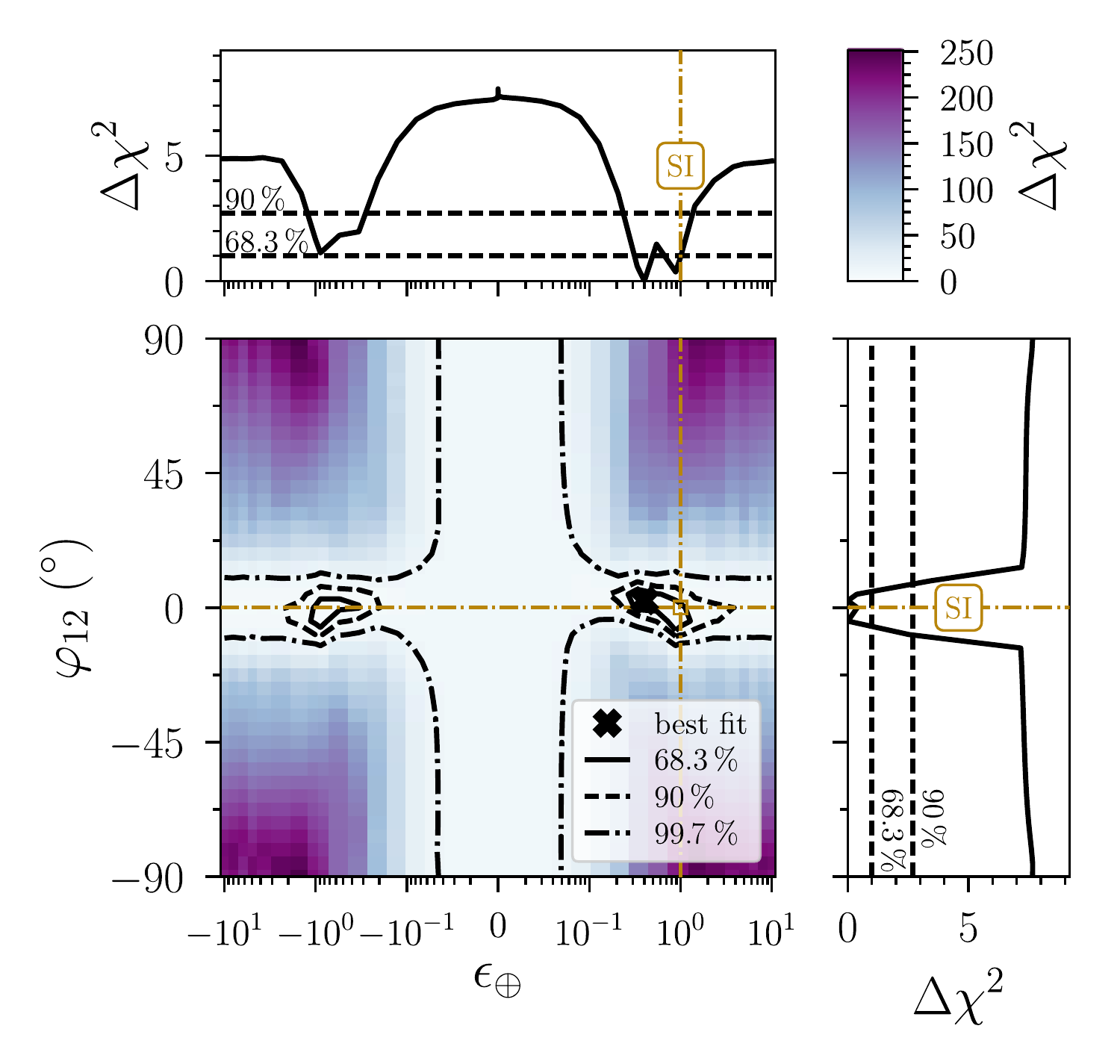}
	\end{minipage}\hfill
	\begin{minipage}[c]{0.5\textwidth}
		\includegraphics[width=.9\linewidth]{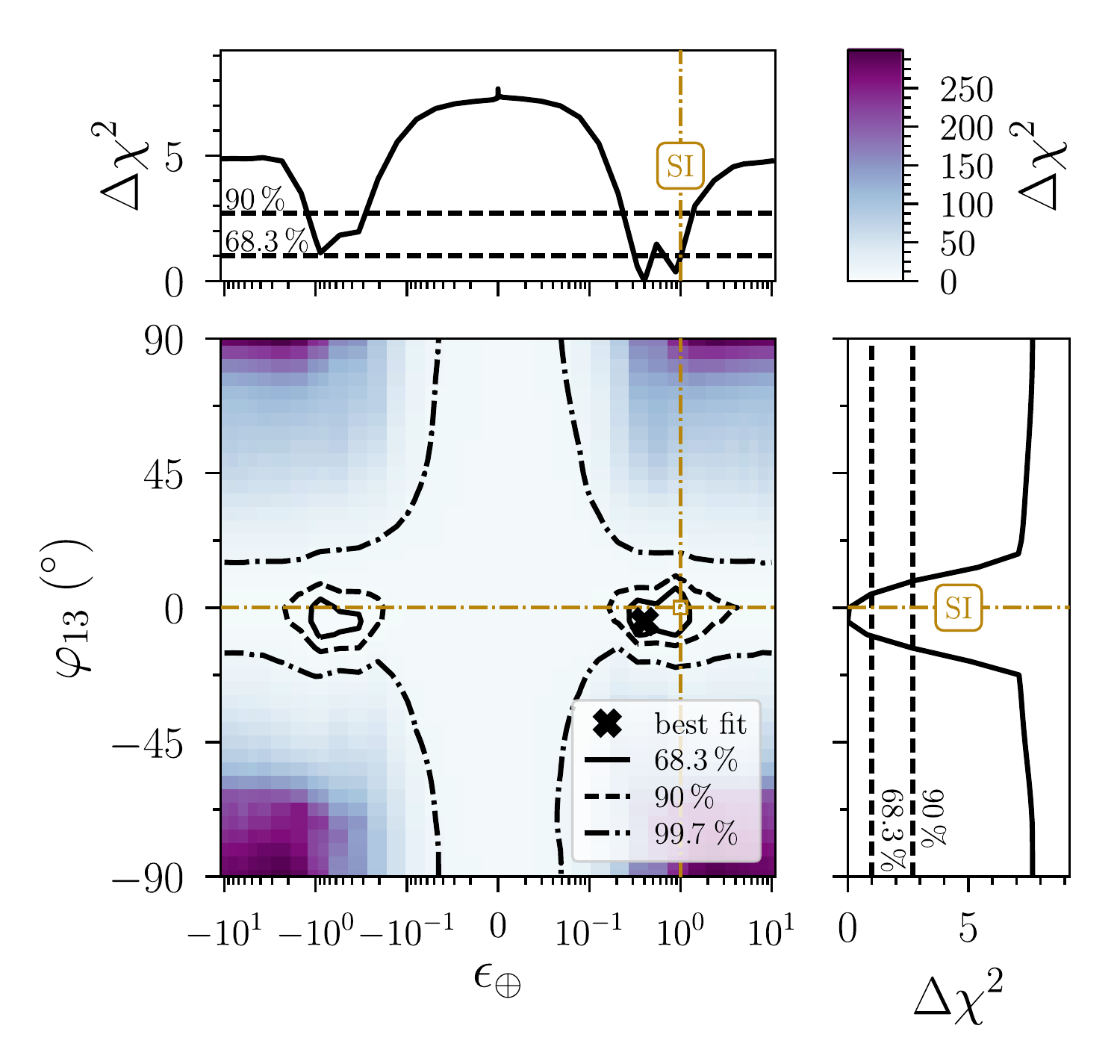}
	\end{minipage}\\
	\begin{minipage}[c]{\textwidth}
	    \centering
		\includegraphics[width=.45\linewidth]{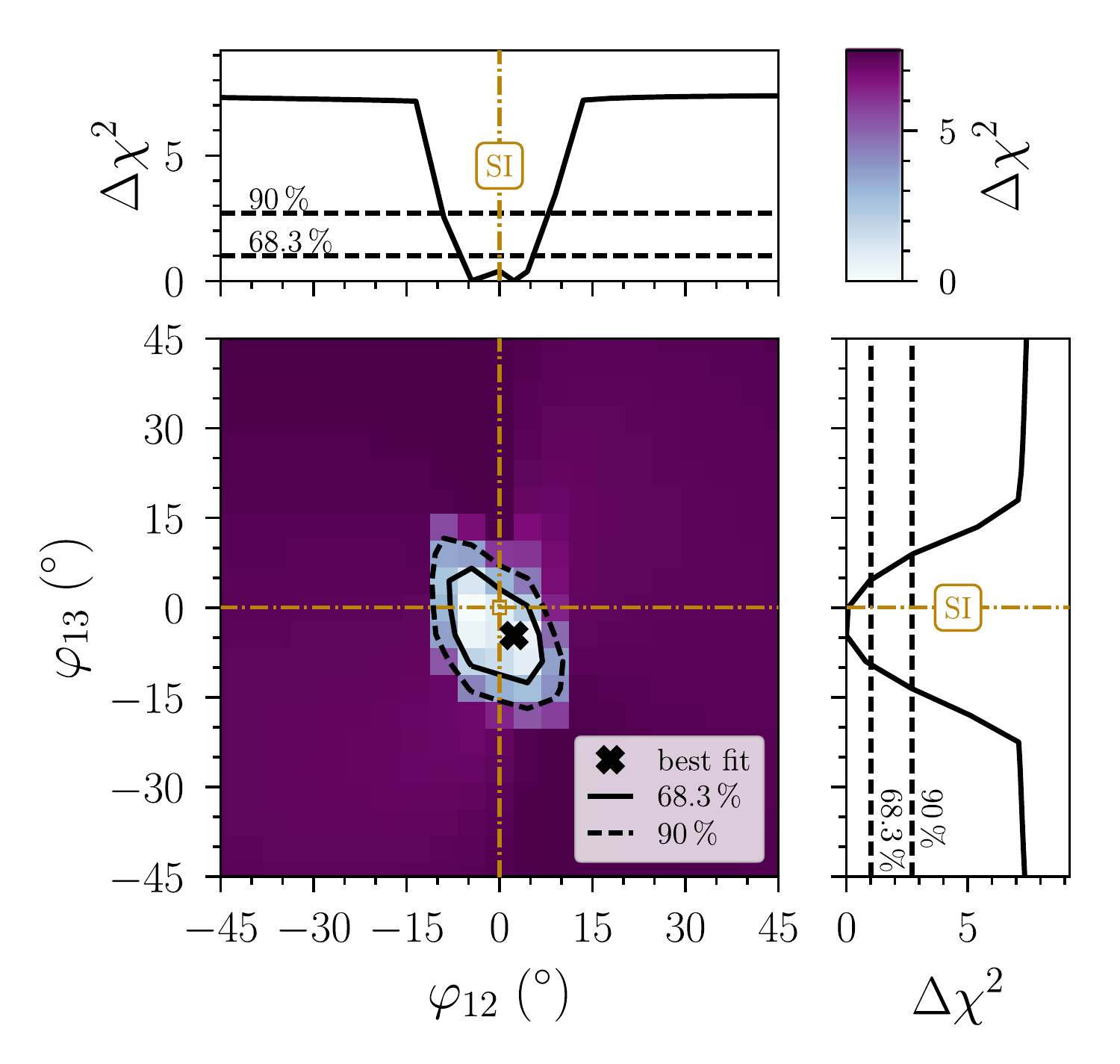}
	\end{minipage}
	\caption{Observed \SI{68.3}{\percent}, \SI{90}{\percent}, and \SI{99.7}{\percent} confidence regions for parameters $\epsilon_\oplus$, $\varphi_{12}$, and $\varphi_{13}$, together with each parameter's projected one-dimensional $\Delta \chi^2$ profile. The color in each of the three large panels encodes the local value of the projected two-dimensional $\Delta \chi^2$ profile. The best fit point for each pair of parameters is indicated by a cross. The SI/flavor-universal NSI hypothesis, indicated by the dash-dotted lines, is located at $\epsilon_\oplus = 1$, $\varphi_{12} = 0$, $\varphi_{13} = 0$. See text for details. This is the first time the GMP overall scale and flavor structure are constrained simultaneously using IceCube DeepCore data.}
	\label{fig:deltachi2 GMP NSI}
\end{figure*}

Our final fit to data employs the generalized matter potential that is characterized by the three intrinsic NSI parameters $(\epsilon_\oplus,\varphi_{12},\varphi_{13})$.
Figure~\ref{fig:deltachi2 GMP NSI} shows the resulting constraints, by means of the projected one- and two-dimensional $\Delta \chi^2$ profiles. In terms of the five standard NSI parameters, the indicated best fit, also given in Table~\ref{tab:result summary}, corresponds to
\begin{align}
    \epsilon^\oplus_{ee} - \epsilon^\oplus_{\mu\mu} = \num{-0.60}\text{ ,}\quad \epsilon^\oplus_{\tau\tau} - \epsilon^\oplus_{\mu\mu} = \num{0.0020}\text{ ,}\nonumber\\
    \epsilon^\oplus_{e\mu} = \num{-0.016}\text{ ,} \quad \epsilon^\oplus_{e\tau} = \num{0.033}\text{ ,} \quad \epsilon^\oplus_{\mu\tau} = \num{-0.0013}\text{ .}\nonumber
    \label{eq:GMP fit in standard parameters}
\end{align}
It is weakly favored over the hypothesis of SI (or flavor-universal NSI) by $\Delta \chi^2 = 2.2$, corresponding to $p = 0.1$, cf. Table~\ref{tab:result summary}. This difference cannot be directly derived from any of the projections in Fig.~\ref{fig:deltachi2 GMP NSI}, as none of them explicitly show the corresponding grid points $(\epsilon_\oplus = \pm 1, \varphi_{12} = 0, \varphi_{13} = 0)$.

The one-dimensional projections yield the following \SI{90}{\percent} confidence intervals (optimized over the two remaining matter potential parameters and all nuisance parameters in each case): $\SI{-9}{\degree} \leq \varphi_{12} \leq \SI{8}{\degree}$, $\SI{-14}{\degree} \leq \varphi_{13} \leq \SI{9}{\degree}$, and the union of intervals \mbox{$\left[-1.2, -0.3\right] \cup \left[0.2, 1.4\right]$} for $\epsilon_\oplus$. The fact that $\varphi_{12}$ and $\varphi_{13}$ are allowed to vary does not have a significant weakening effect on the bounds on $\epsilon_\oplus$ at \SI{90}{\percent} CL, nor does it change the overall shape of its $\Delta \chi^2$ profile (compare $\epsilon^\oplus_{ee} - \epsilon^\oplus_{\mu\mu}$ in Fig.~\ref{fig:deltachi2 NU NSI}). 
The two-dimensional $\Delta \chi^2$ projection onto $(\epsilon_\oplus, \varphi_{12})$ demonstrates that $\abs{\epsilon_\oplus} \geq 0.05$ is excluded at a significance greater than \SI{99.7}{\percent} when $\abs{\varphi_{12}} \geq \SI{10}{\degree}$, for any value of $\varphi_{13}$. Similarly, the projection onto $(\epsilon_\oplus, \varphi_{13})$ implies that $\abs{\epsilon_\oplus} \geq 0.1$ is excluded at a significance greater than \SI{99.7}{\percent} when $\abs{\varphi_{13}} \geq \SI{20}{\degree}$, for any value of $\varphi_{12}$. For smaller values of $\abs{\varphi_{12}}$ and $\abs{\varphi_{13}}$, no \SI{99.7}{\percent} bound on $\epsilon_\oplus$ is obtained.

Conversely, in the projection onto $(\varphi_{12}, \varphi_{13})$, constraints can be set at \SI{90}{\percent} CL. However, the maximal significance of excluding any particular pair of values of the matter rotation angles cannot exceed the $\Delta \chi^2$ value of the vacuum hypothesis, which renders both parameters unphysical. Combined with the lacking bound on $\epsilon_\oplus$ at the \SI{99.7}{\percent} CL this results in the ``crosslike'' shape formed by the corresponding contours in the two upper $\Delta \chi^2$ projections in Fig.~\ref{fig:deltachi2 GMP NSI}.

Finally, we point out that these constraints do not suffer from a loss of generality due to the normal ordering assumption in the fit, for the reasons given in Sec.~\ref{subsec:evolution}.

\section{\label{sec:conclusion}Conclusion}
We have presented a comprehensive study of nonstandard interactions in the propagation of atmospheric neutrinos observed with IceCube DeepCore within the general framework of three flavor neutrino oscillations. Instead of exclusively focusing on NSI in the $\mu$-$\tau$ sector, as was done in our previous analysis~\cite{Aartsen:2017xtt}, we have taken an extended approach that tests all five effective flavor-nonuniversal and flavor-violating NSI coupling strengths for Earth matter individually. In particular, this includes studies of NSI involving the electron flavor, which are not common targets of atmospheric neutrino experiments. All our measurements yield results that are statistically compatible with SM neutrino interactions, i.e., neutrino oscillations with standard matter effects.

The sample of 47855 events with reconstructed energies between \SI{5.6}{\giga\electronvolt} and \SI{100}{\giga\electronvolt} was created from three years of data taken with IceCube DeepCore and contains significant contributions from the interactions of neutrinos and antineutrinos of all flavors. One-by-one NSI parameter fits to this sample result in limits (quoted at \SI{90}{\percent} CL) of similar power with respect to existing global limits on the magnitudes of all five NSI parameters observable by atmospheric neutrino experiments.
Those that apply to $\mu$-$\tau$ nonuniversality and flavor-violation strengths are of the order of $10^{-2}$ and are as, or more, stringent than limits obtained with other oscillation experiments or other IceCube (DeepCore) event samples. 
Weaker $\order{1}$ constraints apply to $e$-$\mu$ nonuniversality, or, when reinterpreted in terms of SM interactions, to the strength of the Earth's standard matter potential.

With a separate fit we have investigated a more general flavor structure of the Earth's matter potential, in a manner similar to recent global NSI fits~\cite{GonzalezGarcia:2011my,Gonzalez-Garcia:2013usa,Esteban:2018ppq}. The adopted parametrization naturally includes NSI hypotheses that lead to cancellations of the induced matter effects in the survival probabilities of atmospheric muon neutrinos and antineutrinos. Within this framework, we have shown that the event sample allows for simultaneous constraints of the overall strength of the matter potential and its flavor structure at \SI{90}{\percent} CL, whereas no constraint emerges at \SI{99.7}{\percent} CL.

Because of the vanishing momentum transfer in the coherent forward scattering processes that generate the neutrino matter potential, our constraints apply independently of the new physics energy scale responsible for NSI. This distinguishes our measurements from those performed at experiments investigating coherent neutrino scattering, deep inelastic neutrino scattering, or at high-energy colliders.

Future versions of this analysis may profit from enhanced minimization approaches, as the computational limitations of this analysis are closely connected to the challenges of minimizing a high-dimensional parameter space with a large number of local minima.
For upcoming NSI measurements with IceCube and its low-energy extension DeepCore, a significant increase in event statistics and an extended energy range compared to the analysis presented in this paper are expected.
Furthermore, the imminent IceCube Upgrade~\cite{upgrade_icrc:2019} will increase the detection efficiency and improve the reconstruction capabilities for atmospheric neutrinos with respect to DeepCore, and lower the energy threshold to allow high-statistics measurements with $\sim \SI{1}{\giga\electronvolt}$ atmospheric neutrinos. It will thus facilitate the determination of the overall strength of the Earth's matter potential and improve IceCube's ability to distinguish NSI from standard matter effects~\cite{PhysRevD.101.032006}.

\begin{acknowledgements}
The IceCube collaboration acknowledges the significant contributions to this manuscript from Thomas Ehrhardt and Elisa Lohfink.
The authors gratefully acknowledge the support from the following agencies and institutions:
USA {\textendash} U.S. National Science Foundation-Office of Polar Programs,
U.S. National Science Foundation-Physics Division,
U.S. National Science Foundation-EPSCoR,
Wisconsin Alumni Research Foundation,
Center for High Throughput Computing (CHTC) at the University of Wisconsin{\textendash}Madison,
Open Science Grid (OSG),
Extreme Science and Engineering Discovery Environment (XSEDE),
Frontera computing project at the Texas Advanced Computing Center,
U.S. Department of Energy-National Energy Research Scientific Computing Center,
Particle astrophysics research computing center at the University of Maryland,
Institute for Cyber-Enabled Research at Michigan State University,
and Astroparticle physics computational facility at Marquette University;
Belgium {\textendash} Funds for Scientific Research (FRS-FNRS and FWO),
FWO Odysseus and Big Science programmes,
and Belgian Federal Science Policy Office (Belspo);
Germany {\textendash} Bundesministerium f{\"u}r Bildung und Forschung (BMBF),
Deutsche Forschungsgemeinschaft (DFG),
Helmholtz Alliance for Astroparticle Physics (HAP),
Initiative and Networking Fund of the Helmholtz Association,
Deutsches Elektronen Synchrotron (DESY),
and High Performance Computing cluster of the RWTH Aachen;
Sweden {\textendash} Swedish Research Council,
Swedish Polar Research Secretariat,
Swedish National Infrastructure for Computing (SNIC),
and Knut and Alice Wallenberg Foundation;
Australia {\textendash} Australian Research Council;
Canada {\textendash} Natural Sciences and Engineering Research Council of Canada,
Calcul Qu{\'e}bec, Compute Ontario, Canada Foundation for Innovation, WestGrid, and Compute Canada;
Denmark {\textendash} Villum Fonden and Carlsberg Foundation;
New Zealand {\textendash} Marsden Fund;
Japan {\textendash} Japan Society for Promotion of Science (JSPS)
and Institute for Global Prominent Research (IGPR) of Chiba University;
Korea {\textendash} National Research Foundation of Korea (NRF);
Switzerland {\textendash} Swiss National Science Foundation (SNSF);
United Kingdom {\textendash} Department of Physics, University of Oxford.
This work has been supported by the Cluster of Excellence ``Precision Physics, Fundamental Interactions, and Structure of Matter'' (PRISMA+ EXC 2118/1) funded by the German Research Foundation (DFG) within the German Excellence Strategy (Project ID 39083149). Parts of this research were conducted using the supercomputer Mogon and/or advisory services offered by Johannes Gutenberg University Mainz (hpc.uni-mainz.de), which is a member of the AHRP (Alliance for High Performance Computing in Rhineland Palatinate,  www.ahrp.info) and the Gauss Alliance e.V.
\end{acknowledgements}

\appendix

\section{\label{app:detailed nsi params}GMP parametrization}

As described in Sec.~\ref{subsec:evolution}, the alternative NSI parametrization that this analysis uses constitutes three rotations: Two real rotations through the angles $\varphi_{12}$ and $\varphi_{13}$ in the $1$-$2$ and $1$-$3$ planes, respectively as well as a complex rotation through the angle $\varphi_{23}$ and the phase $\delta_\mathrm{NS}$. 

This parametrization has the advantage that physics-related arguments allow reducing its dimensionality while approximately retaining model independence. Specifically, Refs.~\cite{Friedland:2004ah,Friedland:2005vy} show that the disappearance of atmospheric muon neutrinos with energies $E_\nu \gtrsim \order{\SI{10}{\giga\electronvolt}}$ proceeds with the same dependence on the baseline-to-energy ratio $L/E_\nu$ as in vacuum when $H_\mathrm{mat}$ has two degenerate eigenvalues. For experiments sensitive mostly to muon neutrino disappearance, this scenario is expected to result in the weakest constraints on matter effects~\cite{Friedland:2004ah,Friedland:2005vy}, and can be realized by setting $\epsilon^\prime_\oplus=0$ in Eq.~(\ref{eq:matter potential eigenvalues}). This in turn renders $\varphi_{23}$ and $\delta_\mathrm{NS}$ unphysical. Such an approach was taken by Refs.~\cite{GonzalezGarcia:2011my,Gonzalez-Garcia:2013usa,Esteban:2018ppq} in their analyses of atmospheric neutrino data. In addition, as argued in~\cite{Esteban:2018ppq}, existing data from atmospheric neutrino experiments has little sensitivity to $CP$-violating effects, which justifies setting the phases $\alpha_{1,2} = 0$. In this case, $H_\mathrm{mat}$ is real and has three parameters, $(\epsilon_\oplus, \varphi_{12}, \varphi_{13})$.

In the remainder of this paper, we refer to this parametrization as the ``generalized matter potential'' (GMP). Any given point in the corresponding three-dimensional parameter space uniquely determines the NSI nonuniversality and flavor-violation parameters in the standard parametrization (see for example~\cite{Esteban:2018ppq}):

\begin{align}
    \epsilon^\oplus_{ee} - \epsilon^\oplus_{\mu\mu} &=& \epsilon_\oplus (\mathrm{cos}^2\phi_{12} - \mathrm{sin}^2\phi_{12})\mathrm{cos}^2\phi_{13} - 1\text{ ,}\\
    \epsilon^\oplus_{\tau\tau} - \epsilon^\oplus_{\mu\mu} &=& \epsilon_\oplus (\sin^2\phi_{13} - \sin^2\phi_{12}\cos^2\phi_{13})\text{ ,}\\
    \epsilon^\oplus_{e\mu} &=& -\epsilon_\oplus \cos\phi_{12} \sin\phi_{12} \cos^2\phi_{13}\text{ ,}\\
    \epsilon^\oplus_{e\tau} &=& -\epsilon_\oplus \cos\phi_{12} \cos\phi_{13} \sin\phi_{13}\text{ ,}\\
    \epsilon^\oplus_{\mu\tau} &=& \epsilon_\oplus \sin\phi_{12} \cos\phi_{13} \sin\phi_{13}\text{ .}
\end{align}\label{eq:eps_ee from vac-like}

When $H_\mathrm{vac}$ is included in the $CP$-conserving framework by setting $\delta_\mathrm{CP}=0$, it is possible to retain the usual minimal parameter ranges for the standard PMNS mixing parameters and neutrino mass-squared differences~\cite{deGouvea:2008nm} by choosing the ranges of the matter-potential rotation angles as $-\pi/2 \leq \varphi_{ij} \leq \pi/2$.

Neutrino evolution is invariant under $H_\nu \to -(H_\nu)^*$~\cite{GonzalezGarcia:2011my}. In vacuum this implies that the two signs of $\Delta m^2_{31(32)}$ cannot be distinguished as long as the most general ranges for $\theta_{12}$ and $\delta_{CP}$ are retained. This degeneracy is broken in matter with SI. However, it reappears as the ``generalized mass ordering degeneracy''\footnote{The presence of NSI would lead to degeneracies that impede the determination of the remaining fundamental unknowns in the neutrino oscillation sector~\cite{deSalas:2017kay,Capozzi:2018ubv,Nufit4.0}. A well-established instance is the so-called ``generalized mass ordering degeneracy''~\cite{Coloma:2016gei}.} once NSI are introduced, because $H_\mathrm{mat}(x) \to -\left[H_\mathrm{mat}(x)\right]^*$ can be implemented by~\cite{Coloma:2016gei}

\begin{align}
		\big[\epsilon^\oplus_{ee}(x) - \epsilon^\oplus_{\mu\mu}(x)\big] &\to -\big[\epsilon^\oplus_{ee}(x) - \epsilon^\oplus_{\mu\mu}(x)\big] - 2\text{ ,}\nonumber\\
		\big[\epsilon^\oplus_{\tau\tau}(x) - \epsilon^\oplus_{\mu\mu}(x)\big] &\to -\big[\epsilon^\oplus_{\tau\tau}(x) - \epsilon^\oplus_{\mu\mu}(x)\big]\text{ ,}\label{eq:GMOD standard NSI}\\
		\epsilon^\oplus_{\alpha\beta}(x) &\to -\epsilon^{\oplus*}_{\alpha\beta}(x)\quad(\alpha \neq \beta)\nonumber\text{ .}
\end{align}
In the Earth, where the effective NSI couplings have little variation along the neutrino trajectory, cf. Eq.~(\ref{eq:constant Earth effective couplings protons neutrons}), the degeneracy is almost exact. When $H_\mathrm{mat}$ is only described by $(\epsilon_\oplus, \varphi_{12}, \varphi_{13})$, it is therefore sufficient to restrict $\Delta m^2_{31(32)} > 0$ and test both signs of $\epsilon_\oplus$. The two choices $(\epsilon_\oplus = \pm 1, \varphi_{12} = 0, \varphi_{13} = 0)$ correspond to neutrino propagation with SI given the normal ordering (``$+$'') and the inverted ordering (``$-$''), respectively.

\section{\label{app:phenomenology nsi at prob lvl}NSI phenomenology at the probability level}

At the \si{\giga\electronvolt}-scale energies considered here, all transitions involving $\nu_e$ are suppressed in vacuum compared to those not involving $\nu_e$. For energies above a few \si{\giga\electronvolt}, $\nu_e \to \nu_{e,\mu,\tau}$ transitions are driven by the mixing angle $\theta_{13}$ and the ``atmospheric'' mass-squared difference $\Delta m^2_{32}$,
with negligible corrections due to the ``solar'' mass-squared difference $\Delta m^2_{21}$. For $\Delta m^2_{32} > 0$ ---the baseline assumption in the example cases in this paper--- SM matter effects in general lead to an enhancement of the transitions involving $\nu_e$, while a negative matter potential in general leads to their suppression.\footnote{In the absence of NSI, the measurement of the mass ordering depends on whether the matter enhancement occurs for neutrinos or antineutrinos~\cite{Aartsen:2019eht}.}

In Figs.~\ref{fig:SM and NSI osc. probs 1d eps_ee} - \ref{fig:SM and NSI osc. probs 1d phi13} oscillation probabilities $P_{\alpha\beta}$ are shown for different NSI parameters.
As all neutrino flavors are considered in this study, no individual oscillation channel can be singled out \textit{a priori}. However, at the energies considered in this paper, the tau neutrino fluxes generated in the atmosphere are negligible~\cite{Bulmahn:2010pg}, resulting in a restriction to $\alpha \in (e, \mu)$. Also, neutrino absorption is not relevant below the $\si{\tera\electronvolt}$ scale~\cite{Giunti:2007ry}. 
In all cases, in the absence of intrinsic $CP$ violation (i.e., $\delta_{CP} = 0$ or $\delta_{CP} = \pi$ and real NSI coupling strengths) the Earth's symmetric matter potential with respect to the midpoint of any given trajectory, $V(x) = V(L - x)$, implies that $P_{\alpha\beta} = P_{\beta\alpha}$ (apart from negligible short-scale corrections due to $h_\mathrm{prod} \neq -d_\mathrm{det}$)~\cite{Akhmedov:2001kd}.

\subsection{Nonzero single NSI parameters}
\paragraph{\label{app:phenomenology rescaled sm potential}Nonzero $\epsilon_\oplus$ (rescaled SM matter potential)}

\begin{figure*}[ht]
\centering
\includegraphics[width=1.0\linewidth]{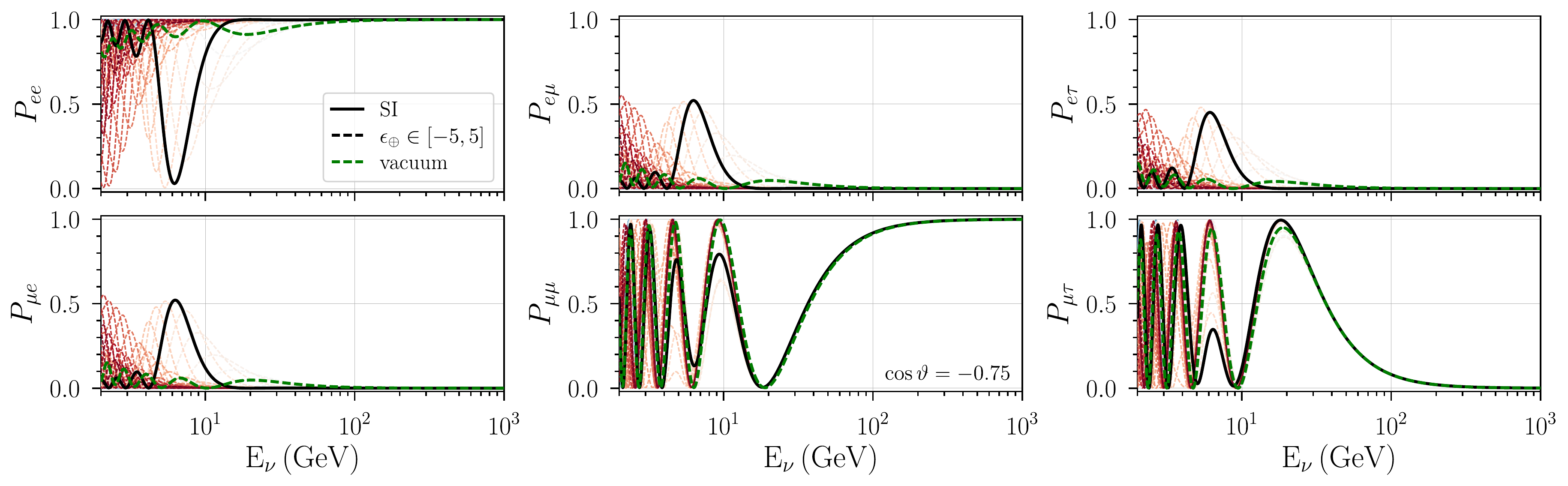}
\caption{\label{fig:SM and NSI osc. probs 1d eps_ee}Oscillation probabilities of atmospheric neutrinos crossing the Earth at zenith angle $\cos(\vartheta)=-0.75$ vs. the neutrino energy $E_\nu$. Shown are different realizations of the effective matter potential strength $\epsilon_\oplus$, with $-5 \leq \epsilon_\oplus \leq 5$. Darker shades represent larger $\abs{\epsilon_\oplus}$. The two cases of SI ($\epsilon_\oplus=1$, in black) and no interactions (vacuum, $\epsilon_\oplus = 0$, in green) are highlighted. The the red dashed lines show those obtained for positive parameter values, the blue dashed lines showing the probabilities obtained for negative values are mostly covered by dark red lines. See text for details.}
\end{figure*}

The oscillation probabilities $P_{\alpha\beta}$ shown in Fig.~\ref{fig:SM and NSI osc. probs 1d eps_ee} result from varying $\epsilon_\oplus \in [-5, 5]$ while restricting the matter potential to the $ee$ entry, i.e., $\varphi_{ij}=0$, yielding effects corresponding to a rescaling of the SM matter potential by the factor $V_\mathrm{CC}(x) \to V^\prime(x) = \epsilon_\oplus V_\mathrm{CC}(x) = \left(1 + \epsilon^\oplus_{ee} - \epsilon^\oplus_{\mu\mu}\right)V_\mathrm{CC}(x)$.

Typically, the $\nu_e$ disappearance probability, $1 - P_{ee}$, in vacuum remains small in the limit $\Delta m^2_{21} \to 0$. In contrast to this, when $\epsilon_\oplus > 0$, the resonance condition can be satisfied (given $\Delta m^2_{32} > 0$ and $\theta_{13} < \pi/4$). In this case, the effective 1-3 mixing angle in matter becomes maximal. A complete disappearance of $\nu_e$ can therefore be observed in principle at a resonance energy $E_R$ which is inversely proportional to the average value of the slowly changing rescaled matter potential along the neutrino trajectory, $\epsilon_\oplus\expval{V_\mathrm{CC}}$. For the Earth's mantle and SI, $E^\mathrm{SI}_R \approx \SI{6}{\giga\electronvolt}$. The example trajectory in Fig.~\ref{fig:SM and NSI osc. probs 1d eps_ee} is chosen such that the oscillation phase leads to a nearly complete disappearance of $\nu_e$ at this energy. The transition probabilities to $\nu_\mu$ and $\nu_\tau$ are nearly identical since the 2-3 mixing is close to maximal. A complete disappearance is also observed for a value of $\epsilon_\oplus \approx 3$ and $E_R \approx \SI{2}{\giga\electronvolt}$. Negative values of $\epsilon_\oplus$ together with $\Delta m^2_{32} > 0$ do not give rise to a similar enhancement. Consequently, there are no significant transitions $\nu_e \to \nu_{\mu,\tau}$ for $\epsilon_\oplus < 0$. Instead, the antineutrino transitions $\bar{\nu}_e \to \bar{\nu}_{e,\mu,\tau}$ are then subject to the matter effects detailed above.

Figure~\ref{fig:SM and NSI osc. probs 1d eps_ee} further demonstrates that $\nu_e$ decouples from the evolution and that the transitions $\nu_{\mu} \to \nu_{\mu,\tau}$ proceed as in vacuum for sufficiently high energy, $E_\nu \gtrsim \SI{20}{\giga\electronvolt}$ for the considered trajectory---irrespective of the value of $\epsilon_\oplus$. At sufficiently low energies of a few \si{\giga\electronvolt}, the simple two-neutrino picture no longer applies, and rather complex corrections due to 1-3 mixing need to be taken into account. For their discussion see, for example, Refs.~\cite{Akhmedov:2006hb,Blennow:2013rca}.

\paragraph{\label{app:phenomenology mu-tau NSI}Nonzero $\epsilon^\oplus_{\tau\tau} - \epsilon^\oplus_{\mu\mu}$ or $\epsilon^\oplus_{\mu\tau}$}

\begin{figure*}[t]
\centering
\includegraphics[width=1.0\linewidth]{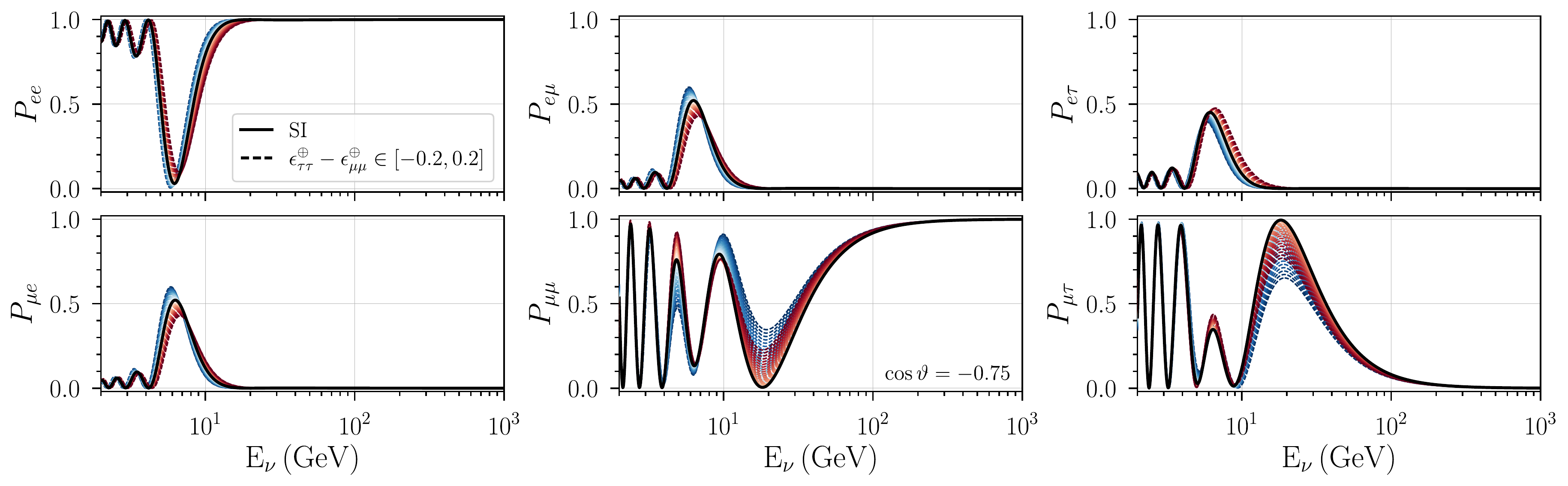}
\caption{\label{fig:SM and NSI osc. probs 1d eps_tautau}Same as Fig.~\ref{fig:SM and NSI osc. probs 1d eps_ee}, but for different realizations of the NSI nonuniversality strength $\epsilon^\oplus_{\tau\tau} - \epsilon^\oplus_{\mu\mu}$, with $-0.20 \leq \epsilon^\oplus_{\tau\tau} - \epsilon^\oplus_{\mu\mu} \leq 0.20$. The blue dashed lines show the probabilities obtained for $\epsilon^\oplus_{\tau\tau} - \epsilon^\oplus_{\mu\mu} < 0$, while the red dashed lines show those obtained for $\epsilon^\oplus_{\tau\tau} - \epsilon^\oplus_{\mu\mu} > 0$. Darker shades represent larger $\abs{\epsilon^\oplus_{\tau\tau} - \epsilon^\oplus_{\mu\mu}}$. See text for details.}
\end{figure*}

In case $\epsilon^\oplus_{\tau\tau} - \epsilon^\oplus_{\mu\mu}$ is the only source of NSI---cf. Fig.~\ref{fig:SM and NSI osc. probs 1d eps_tautau} for variations $\epsilon^\oplus_{\tau\tau} - \epsilon^\oplus_{\mu\mu} \in [-0.20, 0.20]$---the nonuniversality gives rise to an effective potential in the decoupled $\mu$-$\tau$ system that was introduced in the previous section.
As detailed in~\cite{Esmaili:2013fva}, the 2-3 mixing in matter is modified according to the standard MSW mechanism, but with a potential $V^\prime(x) = \left(\epsilon^\oplus_{\tau\tau} - \epsilon^\oplus_{\mu\mu}\right)V_\mathrm{CC}(x)$. For a given sign of the nonuniversality, whether the resonance occurs in the neutrino or the antineutrino channel depends on the octant of $\theta_{23}$. Since the 2-3 mixing in vacuum is nearly maximal, the introduction of nonzero $\epsilon^\oplus_{\tau\tau}-\epsilon^\oplus_{\mu\mu}$ in general leads to a reduction of the mixing. The main observable consequence of $\epsilon^\oplus_{\tau\tau}-\epsilon^\oplus_{\mu\mu}$ is therefore the increased survival probability of both atmospheric $\nu_\mu$'s and $\bar{\nu}_\mu$'s across the broad range of energies at which the $\mu$-$\tau$ system is decoupled. 

\begin{figure*}[t]
\centering
\includegraphics[width=1.0\linewidth]{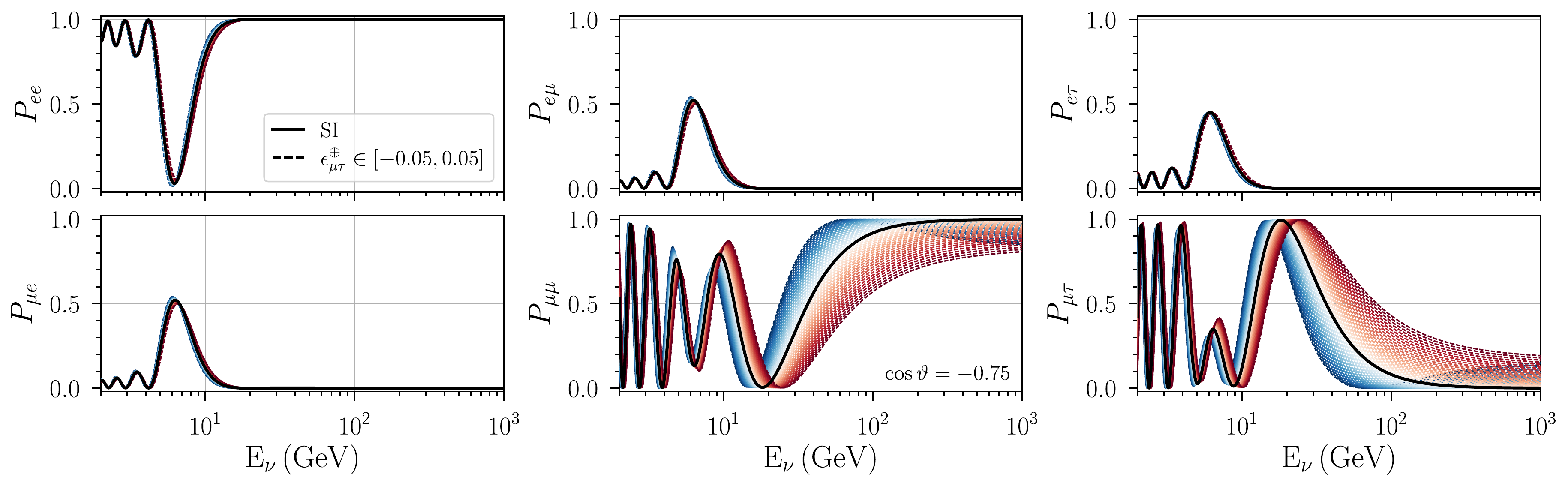}
\caption{\label{fig:SM and NSI osc. probs 1d eps_mutau}Same as Fig.~\ref{fig:SM and NSI osc. probs 1d eps_ee}, but for different realizations of the NSI coupling strength $\epsilon^\oplus_{\mu\tau}$, with $-0.05 \leq \epsilon^\oplus_{\mu\tau} \leq 0.05$. The blue dashed lines show the probabilities obtained for $\epsilon^\oplus_{\mu\tau} < 0$, while the red dashed lines show those obtained for $\epsilon^\oplus_{\mu\tau} > 0$. Darker shades represent larger $\abs{\epsilon^\oplus_{\mu\tau}}$. See text for details.}
\end{figure*}

In contrast to the $\epsilon^\oplus_{\tau\tau} - \epsilon^\oplus_{\mu\mu}$-only case, when $\epsilon^\oplus_{\mu\tau}$ is the only nonzero NSI coupling strength---cf. Fig.~\ref{fig:SM and NSI osc. probs 1d eps_mutau}---, the off-diagonal elements $V_\mathrm{CC}(x) \epsilon^{\oplus(*)}_{\mu\tau}$ of the two-neutrino interaction Hamiltonian result in qualitatively different effects on the neutrino evolution~\cite{Esmaili:2013fva}. 
A resonance occurs for neutrinos when $\epsilon^\oplus_{\mu\tau} < 0$ and for antineutrinos when $\epsilon^\oplus_{\mu\tau} > 0$, independent of the octant of $\theta_{23}$.  Resonances at $E_R \gtrsim \SI{60}{\giga\electronvolt}$ are observed. Since the corresponding oscillation phases are small, the survival probability $P_{\mu\mu}$ becomes nearly maximally enhanced at high energies when $\epsilon^\oplus_{\mu\tau} < 0$. At energies sufficiently far below the resonance, $\epsilon^\oplus_{\mu\tau}$ results in a shift in energy of the oscillation pattern in the $\mu$-$\tau$ system~\cite{Esmaili:2013fva}. When $\epsilon^\oplus_{\mu\tau} > 0$, a shift to higher energies appears for neutrinos, and a shift to lower energies for antineutrinos; the effects are reversed for $\epsilon^\oplus_{\mu\tau} < 0$. At high energies, the two-neutrino survival probability of both $\nu_\mu$ and $\bar{\nu}_\mu$ is reduced compared to the vacuum value~\cite{Esmaili:2013fva}.

\paragraph{\label{subsec:phenomenology FC electron NSI}Subdominant single NSI parameters: $\epsilon^\oplus_{e\mu}$ and $\epsilon^\oplus_{e\tau}$}

\begin{figure*}[t]
\centering
\includegraphics[width=1.0\linewidth]{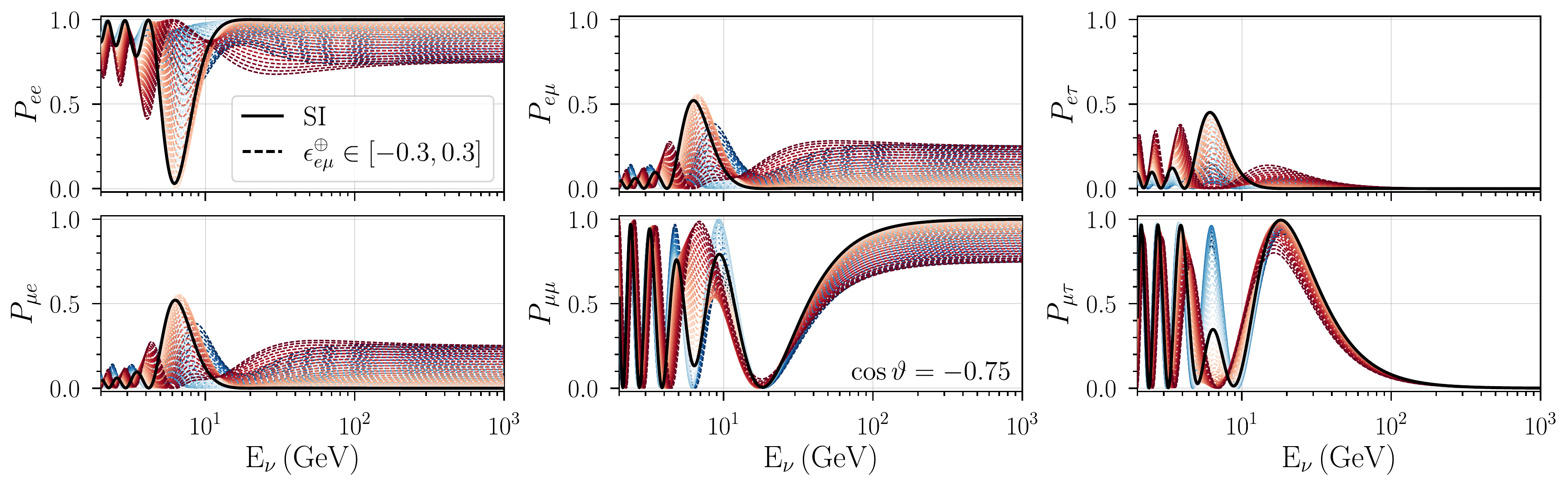}
\caption{\label{fig:SM and NSI osc. probs 1d eps_emu}Same as Fig.~\ref{fig:SM and NSI osc. probs 1d eps_ee}, but for different realizations of the NSI coupling strength $\epsilon^\oplus_{e\mu}$, with $-0.30 \leq \epsilon^\oplus_{e\mu} \leq 0.30$. The blue dashed lines show the probabilities obtained for $\epsilon^\oplus_{e\mu} < 0$, while the red dashed lines show those obtained for $\epsilon^\oplus_{e\mu} > 0$. Darker shades represent larger $\abs{\epsilon^\oplus_{e\mu}}$. See text for details.}
\end{figure*}

Similar to $\epsilon^\oplus_{ee} - \epsilon^\oplus_{\mu\mu}$, the flavor-violating couplings involving the electron flavor, $\epsilon^\oplus_{e\mu}$ and $\epsilon^\oplus_{e\tau}$, typically are not in the focus of atmospheric neutrino studies, partly due to their weaker impact on the disappearance probabilities of $\overset{\brabarb}{\nu}\vphantom{\nu}_\mu$: It has been shown perturbatively that flavor-violating couplings involving the electron flavor contribute to disappearance probabilities of $\overset{\brabarb}{\nu}\vphantom{\nu}_\mu$ only at second order, far away from the 1-3 MSW resonance regime~\cite{PhysRevD.77.013007}. They enter the oscillation probabilities involving the electron flavor at the second order, lower by one order compared to the four remaining couplings~\cite{Kikuchi:2008vq}.

The range of oscillation probabilities induced by values of $\epsilon^\oplus_{e\mu} \in [-0.30, 0.30]$ are depicted in Fig.~\ref{fig:SM and NSI osc. probs 1d eps_emu}. One prominent effect of this is the suppression of the trajectory dependent SI $\nu_e$ resonance around $E_\nu \gtrsim \SI{6}{\giga\electronvolt}$ for large absolute coupling values. 
Varying $\epsilon^\oplus_{e\tau}$ within the same range as $\epsilon^\oplus_{e\mu}$ results in very similar patterns given the exchange of the flavor indices $\mu \leftrightarrow \tau$~\cite{Kikuchi:2008vq}. Hence, only $\epsilon^\oplus_{e\mu}$ results in modifications of the atmospheric oscillation channels involving $\nu_\mu$ across the full range of energies. Characteristically, at the energies shown here it manifests itself in the disappearance of $\nu_\mu$ and $\bar{\nu}_\mu$ and the appearance of $\nu_e$ and $\bar{\nu}_e$. In contrast, $\epsilon^\oplus_{e\tau}$ induces the conversion $\overset{\brabarb}{\nu}\vphantom{\nu}_e \leftrightarrow \overset{\brabarb}{\nu}\vphantom{\nu}_\tau$ at high energies.

For detailed phenomenological and numerical discussions of the oscillation-probability impact of $\epsilon^\oplus_{e\mu}$ and $\epsilon^\oplus_{e\tau}$ at the \si{\giga\electronvolt} energy scale (in the context of future long-baseline and atmospheric neutrino experiments) see, e.g., Refs.~\cite{PhysRevD.88.013001,PhysRevD.93.093017,PhysRevD.100.115034}.

\subsection{\label{subsec:phenomenology arbitrary NSI flavour struct}Arbitrary NSI flavor structure}
In the generalized matter potential parametrized by Eq.~(\ref{eq:vacuum Hamiltonian like parameterisation}), $\epsilon_\oplus$ plays the role of the strength of the matter potential and drives the overall sizes of the coupling strengths. This is evident from the fact that all elements of $H_\mathrm{mat}$ are $\propto \epsilon_\oplus$, cf. Appendix~\ref{app:detailed nsi params}. 

\begin{figure*}[t]
\centering
\includegraphics[width=1.0\linewidth]{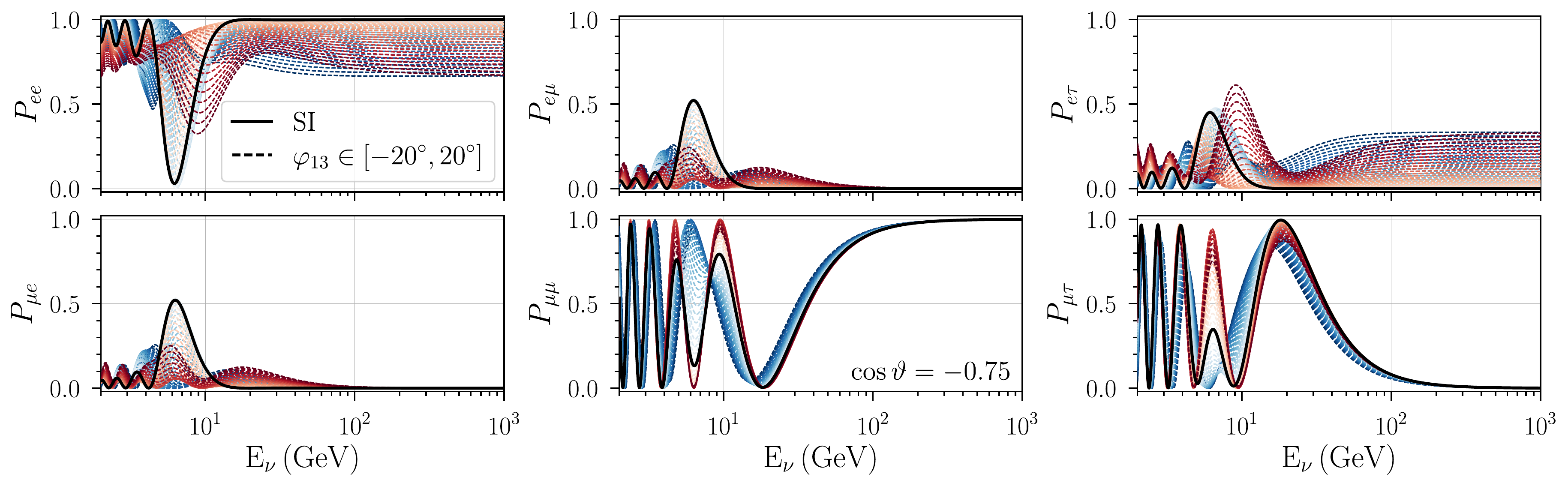}
\caption{\label{fig:SM and NSI osc. probs 1d phi13}Same as Fig.~\ref{fig:SM and NSI osc. probs 1d eps_ee}, but for different realizations of the matter rotation angle $\varphi_{13}$, with $-\SI{20}{\degree} \leq \varphi_{13} \leq \SI{20}{\degree}$, keeping $\epsilon_\oplus = 0$ and $\varphi_{12} = \num{0}$ fixed. The blue dashed lines show the probabilities obtained for $\varphi_{13} < \num{0}$, while the red dashed lines show those obtained for $\varphi_{13} > \num{0}$. Darker shades represent larger $\abs{\varphi_{13}}$. See text for details.}
\end{figure*}

Once $H_\mathrm{mat}$ is allowed to take an arbitrary flavor structure, atmospheric neutrino oscillation probabilities are not, in general, treatable analytically, prompting the implementation of well motivated constraints on the parameter space to yield a point of reference for a phenomenological discussion.
As discussed in Refs.~\cite{Friedland:2004ah,Friedland:2005vy,PhysRevD.78.093002}, in specific regimes of neutrino propagation the three-neutrino evolution in the presence of NSI can be reduced to an analytically treatable effective two-neutrino system, which is rotated with respect to the flavor basis. 
The specific case investigated is when all NSI parameters are zero except $\epsilon^\oplus_{ee} - \epsilon^\oplus_{\mu\mu}$, $\epsilon^\oplus_{\tau\tau} - \epsilon^\oplus_{\mu\mu}$, and $\epsilon^\oplus_{e\tau}$. Here, two identical eigenvalues result in the ``atmospheric parabola'' relation
\begin{equation}
    \epsilon^\oplus_{\tau\tau} - \epsilon^\oplus_{\mu\mu} = \frac{\abs{\epsilon^\oplus_{e\tau}}^2}{1 + \epsilon^\oplus_{ee} - \epsilon^\oplus_{\mu\mu}}\text{ ,}
    \label{eq:atmospheric parabola}
\end{equation}
which is able to accommodate two-flavor vacuumlike (cf. Sec.~\ref{subsec:evolution}) $\overset{\brabarb}{\nu}\vphantom{\nu}_\mu$ disappearance at high energy, independent of the overall sizes of the involved NSI coupling strengths.
The relations given in Appendix~\ref{app:detailed nsi params} demonstrate that Eq.~(\ref{eq:atmospheric parabola}) is satisfied for $\varphi_{12} = 0$, in which case the flavor-violating coupling strengths involving the $\mu$ flavor are zero, $\epsilon^\oplus_{e\mu} = \epsilon^\oplus_{\mu\tau} = 0$.

Hence, as a point of reference, we show the oscillation probabilities obtained for different values of $\varphi_{13} \in [-\SI{20}{\degree},\SI{20}{\degree}]$ while keeping the overall strength of the matter potential and the 1-2 matter rotation angle fixed at $\epsilon_\oplus = 1$ and $\varphi_{12} = 0$, respectively, in Fig.~\ref{fig:SM and NSI osc. probs 1d phi13}. 
In contrast to the behavior resulting from this, the case $\varphi_{13} = 0$ and $\varphi_{12} \neq 0$ results in high-energy $\overset{\brabarb}{\nu}\vphantom{\nu}_e \leftrightarrow \overset{\brabarb}{\nu}\vphantom{\nu}_\mu$ transitions (not shown).

\section{\label{app:mod chi2}Modified Pearson's \texorpdfstring{$\chi^2$}{X²}}

The modified Pearson's $\chi^2$ function used here is defined as~\cite{Aartsen:2017nmd,Aartsen:2019tjl}
\begin{equation}
    \chi^2 = \sum_{i=1}^{N_\mathrm{bins}} \frac{\left(n^\mathrm{obs}_i - n^\mathrm{exp}_i\right)^2}{n^\mathrm{exp}_i + \left(\sigma^\mathrm{exp}_i \right)^2} + \sum_{j=1}^{N_\mathrm{prior}}\frac{\left(\Delta p_j\right)^2}{\sigma^2_{p_j}}
    \label{eq:chi2 function}\text{ .}
\end{equation}
Here, $n^\mathrm{obs}_i$ is the observed number of events in bin $i$ and $n^\mathrm{exp}_i$ is the combined expectation due to $\overset{\brabarb}{\nu}\vphantom{\nu}$ and background events in the same bin. The expectation $n^\mathrm{exp}_i$ depends on the values of the hypothesis parameters of interest and on the values of several nuisance parameters (cf. Sec.~\ref{subsec:nuisance}). Its $\overset{\brabarb}{\nu}\vphantom{\nu}$ contribution is retrieved by reweighting a large sample of simulated events, with an effective livetime that exceeds that of the observed event sample by one order of magnitude.
The variance of the expectation, $\left(\sigma^\mathrm{exp}_i\right)^2$, is given by the sum
\begin{equation}
    \left(\sigma^\mathrm{exp}_i\right)^2 = \sigma^2_{i,\nu} + \sigma^2_{i,\mathrm{bkg}}\label{eq:expectation variance}
\end{equation}
of the variance $\sigma^2_{i, \nu}$ of the expected number of $\overset{\brabarb}{\nu}\vphantom{\nu}$ events and the variance $\sigma^2_{i,\mathrm{bkg}}$ of the expected number of background events in the bin (cf. Table~\ref{tab:event counts}). Generally, the two variances on the right hand side of Eq.~(\ref{eq:expectation variance}) are found to be of similar size, except for some bins in the downgoing region with $\cos(\vartheta_\mathrm{reco}) > 0.5$, where the uncertainty of the background expectation dominates. This is also the only region of the histogram for which the total variance of the expectation, $(\sigma^\mathrm{exp}_i)^2$, can reach a similar size as the Poisson variance $n^\mathrm{exp}_i$. Sensitivity to NSI on the contrary predominantly originates from the upgoing region, $\cos(\vartheta_\mathrm{reco}) < 0$.

The second sum contributing to $\chi^2$ in Eq.~(\ref{eq:chi2 function}) is taken over only those $N_\mathrm{prior}$ nuisance parameters that are subject to external Gaussian constraints: a deviation $\Delta p_j$ of the $j$th such parameter from its nominal value is penalized according to the parameter's prior standard deviation $\sigma_{p_j}$ as $\left(\Delta p_j\right)^2/\sigma^2_{p_j}$.

\begin{figure*}[th!]
\centering
\includegraphics[width=\linewidth]{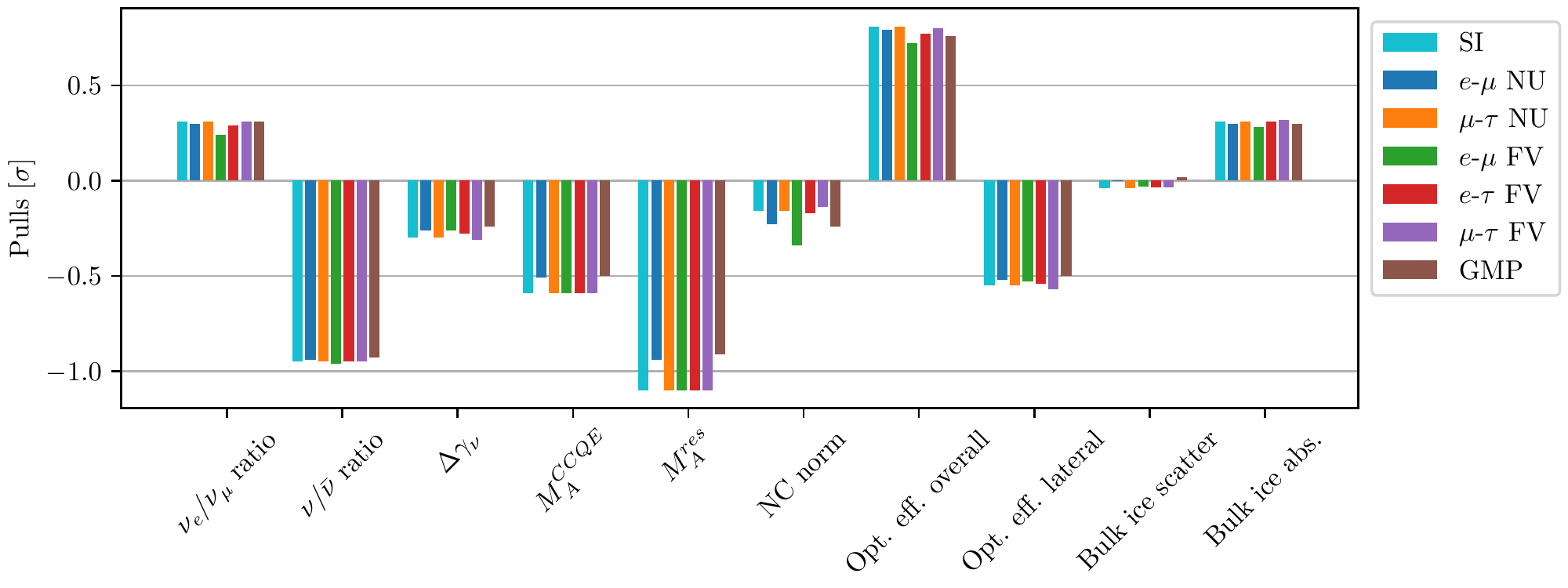}
\caption{\label{fig:pulls}Statistical pulls on the best fit values for all nuisance parameters to which Gaussian priors are associated (cf. Table~\ref{tab:nuisance parameters}), shown for each of the fit hypotheses listed in Table~\ref{tab:NSI models}. More detail can be found in the text.}
\end{figure*}

\section{\label{app:systematics}Nuisance parameter pulls}

Ten of the 15 nuisance parameters that are optimized together with each considered set of NSI parameters have a Gaussian prior associated, as was introduced in Sec.~\ref{subsec:nuisance}. The statistical pulls on the best fit values of these nuisance parameters show little variance between the single fit hypotheses (see Table~\ref{tab:NSI models}), showing the small impact on the expected signal of the different best fit NSI parameter hypotheses. In addition, the statistical pulls on the nuisance parameter fit values are within $1.1\sigma$ in all of the fits (see Fig.~\ref{fig:pulls}), which is expected in case of correctly chosen nuisance parameter priors and ranges. All best fit values of nuisance parameters with no Gaussian prior are well within their allowed ranges listed in Table~\ref{tab:nuisance parameters}.

\section{\label{app:detailed eps ee flat sensitivity}Observations in \texorpdfstring{$e$-$\mu$}{e-mu} nonuniversality}
The relatively constant exclusion power observed for the largest probed values of $\abs{\epsilon^\oplus_{ee} - \epsilon^\oplus_{\mu\mu}}$ is the result of a combination of several probability-level and detector effects, which would hold (at a higher overall level of $\Delta \chi^2$) even if neutrinos could be distinguished from antineutrinos and if interactions of different neutrino flavors could be told apart (cf. Sec.~\ref{subsec:NSI in oscillation probs}): for large positive values of $\epsilon^\oplus_{ee} - \epsilon^\oplus_{\mu\mu}$ in the case of neutrinos, the matter resonance in transitions involving $\nu_e$ shifts to below the detection energy threshold, leading to the suppression (compared to the SI scenario) of oscillations at neutrino energies just above the detection threshold. A similar suppression of oscillations occurs for large negative values of the nonuniversality. The summation over neutrinos and antineutrinos, as well as over appearance and disappearance channels, results in a further weakening of the NSI signature. Moreover, near the detection threshold the discrimination power between $\overset{\brabarb}{\nu}\vphantom{\nu}_\mu$ CC events and events of other types is impeded due to the small propagation distance of the $\mu^\pm$ emerging at the interaction vertex of each of the former.

Close to the position of the large peak at $\epsilon^\oplus_{ee} - \epsilon^\oplus_{\mu\mu} \approx -1$, standard matter effects are compensated by NSI, giving rise to vacuum oscillations. These are disfavored by the data at $\Delta \chi^2 \approx 7.2$ with respect to the best fit at $\epsilon^\oplus_{ee} - \epsilon^\oplus_{\mu\mu} \approx -0.59$, somewhat more strongly than expected from the \SI{90}{\percent} sensitivity range. Compared to the hypothesis of $e$-$\mu$ flavor universality (or SI), vacuum oscillations are disfavored at the level of $\Delta \chi^2 \approx 5.9$. Note that vacuum oscillations are not necessarily expected to provide the worst fit to SI in practice, since the neutrino event distributions (of any flavor) disfavor other intervals in $\epsilon^\oplus_{ee} - \epsilon^\oplus_{\mu\mu}$ than do their antineutrino counterparts.


\providecommand{\noopsort}[1]{}\providecommand{\singleletter}[1]{#1}%
%


\end{document}